\documentclass[twocolumn,showpacs,preprintnumbers,amsmath,amssymb]{revtex4}

\usepackage{epsfig}
\usepackage{subfigure}
\usepackage{bm}
\usepackage{graphicx}
\usepackage{dcolumn}
\usepackage{epsfig,amsmath}
\graphicspath{{pics/}}    

\newcommand{\be}{\begin{equation}}
\newcommand{\ee}{\end{equation}}
\newcommand{\bea}{\begin{eqnarray}}
\newcommand{\eea}{\end{eqnarray}}

\newcommand{\nn}{\nonumber\\}

\newcommand{\UNIT}[1]{\mbox{$\,{\rm #1}$}}

\newcommand{\fm}{\UNIT{fm}}

\def\bi{\begin{itemize}}
\def\ei{\end{itemize}}

\def\fm3{$\mathrm{fm}^3$}
\def\f3{\mathrm{fm}^3}

\def\be{\begin{equation}}
\def\ee{\end{equation}}

\def\B0bar{$\bar{B^0}$}

\begin{document}

\title{Hadronic and partonic sources of direct photons in relativistic heavy-ion collisions}

\author{O.~Linnyk}

\email{Olena.Linnyk@theo.physik.uni-giessen.de}

\affiliation{%
 Institut f\"ur Theoretische Physik, %
  Universit\"at Giessen, %
  35392 Giessen, %
  Germany %
}

\author{V.~Konchakovski}%
\affiliation{%
 Institut f\"ur Theoretische Physik, %
  Universit\"at Giessen, %
  35392 Giessen, %
  Germany %
}

\author{T.~Steinert}%
\affiliation{%
 Institut f\"ur Theoretische Physik, %
  Universit\"at Giessen, %
  35392 Giessen, %
  Germany %
}

\author{W.~Cassing}
\affiliation{%
 Institut f\"ur Theoretische Physik, %
  Universit\"at Giessen, %
  35392 Giessen, %
  Germany %
}

\author{E.~L.~Bratkovskaya}%
\affiliation{%
 Institut f\"ur Theoretische Physik, %
 Johann Wolfgang Goethe University, %
 60438 Frankfurt am Main, %
 Germany; %
Frankfurt Institute for Advanced Studies, %
 60438 Frankfurt am Main, %
 Germany; %
}




\date{\today}

\begin{abstract}
The direct photon spectra and flow ($v_2$, $v_3$) in heavy-ion
collisions at SPS, RHIC and LHC energies are investigated within a
relativistic transport approach incorporating both hadronic and
partonic phases -- the Parton-Hadron-String Dynamics (PHSD). In the
present work, four extensions are introduced compared to our
previous calculations: (i) going beyond the soft-photon
approximation (SPA) in the calculation of the bremsstrahlung
processes $meson+meson\to meson+meson+\gamma$, (ii)  quantifying the
suppression due to the Landau-Pomeranchuk-Migdal (LPM) coherence
effect, (iii) adding the additional channels $V+N\to N+\gamma$ and
$\Delta\to N+\gamma$ and (iv) providing PHSD calculations for Pb+Pb
collisions at $\sqrt{s_{NN}}$ = 2.76 TeV. The first issue extends
the applicability of the bremsstrahlung calculations to higher
photon energies in order to understand the relevant sources in the
region $p_T=0.5-1.5$~GeV, while the LPM correction turns out to be
important for $p_T<0.4$~GeV in the partonic phase. The results
suggest that a large elliptic flow $v_2$ of the direct photons
signals a significant contribution of photons produced in
interactions of secondary mesons and baryons in the late (hadronic)
stage of the heavy-ion collision. In order to further differentiate
the origin of the direct photon azimuthal asymmetry (late hadron
interactions vs electromagnetic fields in the initial stage), we
provide predictions for the photon spectra, elliptic flow and
triangular flow $v_3(p_T)$ of direct photons at different
centralities to be tested by the experimental measurements at the
LHC energies. Additionally, we illustrate the magnitude of the
photon production in the partonic and hadronic phases as functions
of  time and  local energy density. Finally, the 'cocktail' method
for an estimation of the background photon elliptic flow, which is
widely used in the experimental works, is supported by the
calculations within the PHSD transport approach.
\end{abstract}

\pacs{25.75.-q, 13.60.Le, 14.40.Lb, 14.65.Dw}

\maketitle

\section{Introduction}

Direct photons are a powerful probe of the quark-gluon plasma (QGP)  as
created in ultra-relativistic nuclear collisions. The photons
interact only electromagnetically and thus escape to the detector
almost undistorted through the dense and strongly-interacting medium. Thus
the photon transverse-momentum spectra and their azimuthal asymmetry
carry  information on the properties of the matter under extreme
conditions, existing in the first few fm/c of the collisional
evolution.

On the other hand, the measured photons provide a time-integrated
picture of the heavy-ion collision dynamics and are emitted from
every moving charge -- partons or hadrons. Therefore, a multitude of
photon sources has to be differentiated in order to access the
signal of interest. The dominant contributions to the inclusive
photon production are the decays of mesons, dominantly pions,
$\eta$- and $\omega$-mesons. Experimental collaborations subtract
the ``decay photons" from the inclusive photon spectrum using a
cocktail calculation~\cite{PHENIX1,Wilde:2012wc} and obtain the
``direct" photons.

In particular the direct photons at transverse momenta ($p_T<$ 3
GeV) are expected to be dominated by  "thermal" sources, i.e. the
radiation from the strongly interacting Quark-Gluon-Plasma
(sQGP)~\cite{Shuryak:1978ij} and the secondary meson+meson and
meson+baryon interactions in the hadronic
phase~\cite{Song:1994zs,Li:1998ma}. These partonic and hadronic
channels have been studied within PHSD in detail in
Refs.~\cite{Linnyk:2013hta,Linnyk:2013wma} at
Relativistic-Heavy-Ion-Collider (RHIC) energies. It was found that
the partonic channels constitute up to  half of the observed direct
photon spectrum for very central collisions. Other theoretical
calculations also find a significant contribution of the photons
produced in the QGP to the direct photon spectrum
\cite{Chatterjee:2005de,Liu:2009kq,Dion:2011vd,Dion:2011pp,Chatterjee:2013naa,Shen:2013vja}.

The low-$p_T$ direct photons probe not only the
temperature~\cite{Shen:2013vja,PHENIX1,Wilde:2012wc} of the produced
QCD-matter, but also its (transport) properties, for instance, the
shear viscosity. Using the direct photon elliptic flow $v_2$ (a
measure of the azimuthal asymmetry in the photon distribution) as a
viscosimeter was first suggested by Dusling et al. in
Ref.~\cite{Dusling:2009bc}; this idea was later supported by the
calculations in Refs.~\cite{Dion:2011vd,Dion:2011pp,
Shen:2013vja,Shen:2013cca}. It was also suggested that the photon
spectra and $v_2$ are sensitive to the collective directed flow of
the system~\cite{vanHees:2014ida,Shen:2014cga}, to the equation of
state~\cite{Goloviznin:2012dy,vanHees:2014ida}, to the possible
production of a
Glasma~\cite{McLerran:2014hza,Monnai:2014kqa,Liu:2012ax}, to the
rate of chemical equilibration in the
QGP~\cite{Almasi:2014pya,Ozvenchuk:2012fn,Ozvenchuk:2012kh} and to
the asymmetry induced by the strong magnetic field (flash) in the
very early stage of the
collision~\cite{Bzdak:2012fr,Tuchin:2014pka,Tuchin:2012mf,Muller:2013ila}.

However, the recent observation by the PHENIX
Collaboration~\cite{PHENIX1} that the elliptic flow $v_2(p_T)$ of
'direct photons' produced in minimum bias Au+Au collisions at
$\sqrt{s_{NN}}=200$~GeV is comparable to that of the produced pions
was a surprise and in contrast to the theoretical expectations and
predictions. Indeed, the photons produced by partonic interactions
in the quark-gluon plasma phase have not been expected to show
considerable flow because they are dominated by the emission in the
initial phase before the elliptic flow fully develops.

In Refs.~\cite{Linnyk:2013hta,Linnyk:2013wma} we have applied the
PHSD approach to photon production in Au+Au collisions at
$\sqrt{s_{NN}}=200$~GeV and studied the transverse momentum spectrum
and the elliptic flow $v_2$ of photons from hadronic and partonic
production channels. The microscopic description of the full
collision evolution is calculated in the covariant off-shell
transport approach PHSD. The degrees-of-freedom in the partonic and
hadronic phases are strongly interacting dynamical quasi-particles
and off-shell hadrons, respectively.

It was found in Refs. \cite{Linnyk:2013hta,Linnyk:2013wma} that the
PHSD calculations reproduce the transverse momentum spectrum of
direct photons as measured by the PHENIX Collaboration in
Refs.~\cite{PHENIXlast,Adare:2008ab}. The centrality dependence of
the thermal photon yield in PHSD was predicted to be $\sim
N_{part}^\alpha$ with the exponent $\alpha=1.5$, which is in good
agreement with the most recent measurement of
$\alpha=1.48\pm0.08\pm0.04$ by the PHENIX
Collaboration~\cite{Adare:2014fwh}. Also, the PHSD  described the
data on the elliptic flow of inclusive {\em and direct} photons at
the top RHIC energy. The strong $v_2$ of direct photons -- which is
comparable to the hadronic $v_2$ -- in PHSD was attributed to
hadronic channels, i.e. to meson binary reactions which are not
subtracted in the data. As sources for photon production, we have
incorporated the interactions of off-shell quarks and gluons in the
strongly interacting quark-gluon plasma (sQGP) ($q+\bar q\to
g+\gamma$ and
 $q(\bar q)+g\to q(\bar q)+\gamma$), the decays of hadrons
($\pi\to\gamma+\gamma$, $\eta\to\gamma+\gamma$,
$\omega\to\pi+\gamma$, $\eta'\to\rho+\gamma$, $\phi\to\eta+\gamma$,
$a_1\to\pi+\gamma$) as well as their interactions
$\pi+\pi\to\rho+\gamma$, $\rho+\pi\to\pi+\gamma$, meson-meson
bremsstrahlung $m+m\to m+m+\gamma$), meson-baryon bremsstrahlung
($m+B\to m+B+\gamma$). In the present work we additionally
incorporate the two-to-two vector meson+nucleon interactions
($V+p\to\gamma+p/n$ and $V+n\to\gamma+p/n$) and the decay of the
$\Delta$-resonance $\Delta\to N\gamma$.

The photon production via bremsstrahlung in meson-meson and
meson-baryon elastic collisions was found to be a very important
source to interpret the data on the direct photon spectra and
elliptic flow simultaneously~\cite{Linnyk:2013hta,Linnyk:2013wma}.
In the previous works~\cite{Linnyk:2013hta,Linnyk:2013wma}, we have
been calculating the photon bremsstrahlung from all elastic
meson-meson and meson-baryon scatterings $m_1 + m_2$, which occur
during the heavy-ion collisions (including $m_i = \pi, \eta, K, \bar
K, K^0, K^*, \bar K^*, K^{*0}, \eta', \omega, \rho, \phi, a_1$). For
the calculation of the bremsstrahlung cross sections we have been
applying the soft-photon approximation (SPA). Therefore the
resulting yield of the bremsstrahlung photons depended on  model
assumptions such as i) the cross section for the meson-meson elastic
scattering (we assumed 10 mb for all meson species), ii) incoherence
of the individual scatterings and iii) the soft-photon approximation
(i.e. low photon energy and low $\sqrt{s}$ of the collision). The
adequacy of the SPA assumption has been studied before in
Ref.~\cite{Eggers:1995jq} and a theoretical uncertainty of up to a
factor of 2 was found.

In view of the importance of questions that can be answered by
 direct photon measurements, we have
improved our implementation of the photon production in the PHSD
transport approach, in particular of the bremsstrahlung photon
production in the meson-meson collisions. In the scope of the
present work, we depart from the assumptions above by i) using a
covariant one-boson-exchange (OBE) chiral model for the $\pi$
interactions, ii) investigating the suppression due to the coherence
of the photon emission with long wavelength (LPM effect) and iii)
deriving and implementing the ``exact" OBE cross sections for photon
bremsstrahlung thus departing from the SPA.

The results of our calculations are compared to the data from the
SPS and RHIC Collaborations to check if the earlier conclusions and
interpretations hold. Additionally, we provide calculations for the
photon production in $Pb+Pb$ collisions at the energy of
$\sqrt{s_{NN}}=2.76$~TeV (spectra, elliptic $v_2$ and triangular
$v_3$ flow of direct photons). The comparison of these calculations
to the future data of the ALICE Collaboration will be of great
interest since the preliminary
data~\cite{Wilde:2012wc,Lohner:2012ct} indicate a significant direct
photon signal at low $p_T$ with a large elliptic flow.

\section{Photon production in heavy-ion collisions within the PHSD approach}
\label{section_phsd}

The {PHSD} model~\cite{CasBrat,BrCa11} is an off-shell transport
approach that consistently describes the full evolution of a
relativistic heavy-ion collision from the initial hard scatterings
and string formation through the dynamical deconfinement phase
transition to the quark-gluon plasma as well as hadronization and
the subsequent interactions in the hadronic phase. The two-particle
correlations {resulting from the finite width of the parton spectral
functions} are taken into account dynamically {in the PHSD} by means
of the {generalized} off-shell transport equations~\cite{Cass_off1}
that go beyond the mean field or Boltzmann
approximation~\cite{Cassing:2008nn}. The transport theoretical
description of quarks and gluons in the PHSD is based on the
Dynamical Quasi-Particle Model~\cite{Cassing:2007nb} (DQPM) for
partons that is {constructed} to reproduce lattice QCD (lQCD)
results for the entropy density, energy density and pressure as
functions of temperature for the quark-gluon plasma in thermodynamic
equilibrium. In the hadronic sector, PHSD is equivalent to the
Hadron-String Dynamics (HSD)
approach~\cite{Cass99,Brat97,Bratkovskaya:2008iq}. For details about
the DQPM model and the off-shell transport approach we refer the
reader to the review Ref.~\cite{Cassing:2008nn}.

We stress that a non-vanishing width in the partonic spectral
functions is the main difference {between} the DQPM {and}
conventional quasiparticle models~\cite{qp1}. Its influence {on the
collision dynamics} is essentially seen in the correlation
functions. For instance, in the stationary limit, the correlation
involving the off-diagonal elements of the energy-momentum tensor
$T^{kl}$ defines the shear viscosity $\eta$ of the
medium~\cite{Peshier:2005pp}. Here a sizeable width is mandatory to
obtain a small ratio of the shear viscosity to entropy density
$\eta/s$, which results in a roughly hydrodynamical evolution of the
partonic system in PHSD \cite{Cass08}. The finite width leads to
two-particle correlations, which are taken into account dynamically
by means of the {\em generalized} off-shell transport
equations~\cite{Cass_off1}, going beyond the mean field
approximation~\cite{Cassing:2008nn,Linnyk:2011ee}. It has been shown
in~\cite{Rauber:2014mca} that the final width (the imaginary part of
the self energy) is demanded by the causality constraint on the
propagator in the theory of the strongly interacting particles as
soon as the interaction leads to a sizable dressing mass squared (real part
of the self energy).

In the past the PHSD approach has provided a consistent description
of the bulk properties of heavy-ion collisions -- rapidity spectra,
transverse mass distributions, azimuthal asymmetries of various
particle species -- from low Super-Proton-Synchrotron (SPS) up to
the LHC energies~\cite{CasBrat,BrCa11,Konchakovski:2014fya}. In the
hadronic sector, PHSD is equivalent to the Hadron-String Dynamics
(HSD) approach~\cite{Cass99,Brat97,Ehehalt}, in which the photon
production at top SPS energies has been investigated before in
Ref.~\cite{Bratkovskaya:2008iq} with an emphasis on the role of
meson-meson interactions. The PHSD approach was also successfully
used for the analysis of penetrating probes, such as
charm~\cite{Linnyk:2008hp,Linnyk:2007zx} and dilepton production
from hadronic and partonic sources at SPS, RHIC and LHC
energies~\cite{Linnyk:2011hz,Linnyk:2011vx}.

Indeed the calculations within the PHSD have reproduced the measured
differential spectra of dileptons from heavy-ion collisions at SPS
and RHIC energies (see Refs.~\cite{Linnyk:2011hz,Linnyk:2011vx}).
Also, it has been checked in Ref.~\cite{Linnyk:2012pu} that the
dilepton production from the QGP constituents -- as incorporated in
the PHSD~\cite{olena2010,Linnyk:2011hz} -- agrees with the dilepton
rate emitted by the thermalized QCD medium as calculated in the lQCD
approach. Indeed, the deconfined state of matter -- created in
heavy-ion collisions at RHIC~\cite{STARQGP,PHENIXQGP,BRAHMS,PHOBOS}
-- was clearly seen in the dilepton yield above invariant masses of
1.2 GeV~\cite{Linnyk:2011hz,Linnyk:2011vx}.

As sources of photon production - on top of the general dynamical
evolution - we consider hadronic~\cite{Turbide:2003si} as well as
partonic~\cite{Feinberg:1976ua,Shuryak:1978ij} interactions. Let us
first describe all the contributions, which consist of the photon
production in the quark and gluon collisions, from the hadronic
decays and the interactions of  intermediate mesons
produced throughout the evolution of the nucleus-nucleus collision.

1) Photons are radiated by  quarks in the interaction with other
quarks and gluons. In this sense, we differentiate two classes of
processes: first the two-to-two reactions
\begin{eqnarray}
q+\bar{q}\rightarrow g+\gamma,\nonumber\\
q/\bar{q}+g\rightarrow q/\bar{q}+\gamma .
\nonumber\label{quark22a}
\end{eqnarray}
The implementation of the photon production by the quark and gluon
interactions in the PHSD is based on the off-shell cross sections
for the interaction of the massive dynamical quasi-particles as
described in~\cite{olena2010,Linnyk:2013hta}.  In addition, photon
production in the bremsstrahlung reactions $q+q/g\to q+q/g+\gamma$
is possible~\cite{Haglin:1992fy}.

2) All colliding hadronic charges (meson, baryons) can also radiate
photons by the bremsstrahlung processes:
\begin{eqnarray}
m+m\to m+m+\gamma\label{mmBr}\\
m+B\to m+B+\gamma .\label{mBbr}
\end{eqnarray}
The processes (\ref{mmBr}) have been studied within the PHSD in
Refs.~\cite{Bratkovskaya:2008iq,Linnyk:2013hta}, while the processes
(\ref{mBbr}) were added in Ref.~\cite{Linnyk:2013wma}. The
implementation of photon bremsstrahlung from hadronic reactions in
transport approaches has been based until now on the 'soft photon'
approximation (SPA). The soft-photon approximation~\cite{Gale87}
relies on the assumption that the radiation from internal lines is
negligible and the strong interaction vertex is on-shell and is
valid only at low energy (and $p_T$) of the produced photon. Since
the relatively high transverse momenta of the direct photons
($p_T=0.5-1.5$~GeV) are most important for a potential understanding
of the "direct photon puzzle", we extend here the accuracy of our
calculations for the photons produced via  the bremsstrahlung
mechanism beyond the applicability of the SPA. The details of the
new model are given below in section~\ref{sect:brems}.

3) Additionally, the photons can be produced in binary $meson+meson$
and $meson+baryon$ collisions. We consider here the direct photon
production in the following $2\to2$ scattering processes: the
meson+meson collisions
\begin{eqnarray}
\pi + \pi \rightarrow \rho + \gamma , \nonumber\\
\pi + \rho \rightarrow \pi + \gamma ,  \label{22}
\end{eqnarray}
and the meson+baryon collisions
\bea V+N &\to& \gamma+N, \label{rhop} \\ \mbox{where } V=\rho, \
\phi, \  \omega,&&\mbox{ and }N=n,p, \nonumber
\end{eqnarray}
 accounting for all possible charge combinations. The
implementation of the reactions (\ref{22}) is the same as described
in Refs.~\cite{Bratkovskaya:2008iq,Linnyk:2013hta}. On the other
hand, the meson+baryon processes (\ref{rhop}) are incorporated here
for the first time in a transport approach. We describe the relevant
cross sections in section~\ref{sect:22}.

4) Photon production in the decays of mesons ($\pi^0, \eta,
\eta^\prime, \omega, \phi, a_1$) and the $\Delta$-resonance, where
the parent hadrons are produced in baryon-baryon ($BB$),
meson-baryon ($mB$) or meson-meson ($mm$) collisions in the course
of the heavy-ion collision. We consider the contributions from the
photon decays of the following mesons:
\begin{eqnarray}
\pi^0 & \to & \gamma+ \gamma,          \nonumber \\
\eta & \to & \gamma + \gamma,         \nonumber\\
\eta^\prime & \to & \rho + \gamma,    \nonumber\\
\omega & \to & \pi^0 + \gamma,        \nonumber\\
\phi& \to& \eta + \gamma,           \nonumber\\
 a_1 & \to  &\pi + \gamma.             \nonumber\\ \Delta & \to &
\gamma+N, \label{l}\end{eqnarray}
The decay probability is calculated according to the corresponding
branching ratios taken from the latest compilation by the Particle
Data Group~\cite{PDG}, updating slightly the values applied in
earlier HSD investigations at SPS
energies~\cite{Bratkovskaya:2008iq}. The broad resonances --
including the $a_1, \rho, \omega$ mesons -- in the initial or final
state are treated in PHSD in line with their (in-medium) spectral
functions as implemented and described in detail in
Ref.~\cite{Bratkovskaya:2008iq}.

The photon production from the mesonic decays represents a
'background' for the search of the direct photons. However, this
background can only partly be fixed by the independent measurements.
One usually uses the 'cocktail' method to estimate the photon decay
spectra and their contribution to the elliptic flow $v_2$, which
relies among others on the $m_T$-scaling assumption for the particle
spectra. We have assumed throughout that the direct photon spectra
do not include the contributions from the $\pi$, $\eta$, $\eta'$ and
$\omega$ decays, because they were subtracted experimentally.
%

\section{Theoretical developments}

The extensions compared to our previous publications on photon
production in heavy-ion collisions within the PHSD model are
described in the following subsection: an improved implementation of
the Bremsstrahlung channel $m+m\to m+m+\gamma$ beyond the SPA is
described in subsection~\ref{sect:brems}, an estimation of the
suppression due to the Landau-Pomeranchuk-Migdal (LPM) effect will
be presented in subsection~\ref{sect:LPM} and  additional  baryonic
processes will be described in  subsection~\ref{sect:22}.

\subsection{Bremsstrahlung $m+m\to m+m+\gamma$ beyond the soft-photon approximation}
\label{sect:brems}

In the present work, we improve the description of the photon
bremsstrahlung in meson+meson scattering by going beyond the
soft-photon approximation~\cite{Low:1958sn}. Since pions are the
dominant meson species in the heavy-ion collisions, we concentrate
here on the description of the bremsstrahlung photon production in
pion+pion collisions.
Indeed, we recall that the pion+pion interactions are most numerous
and provide the dominant source of photons from the meson+meson
bremsstrahlung mechanism. This was shown in the previous
investigation~\cite{Linnyk:2013hta} explicitly by studying the
channel decomposition of the bremsstrahlung photons. Therefore, we
will be able to considerably reduce the theoretical uncertainty in
 modeling the total photon production in the hadron
bremsstrahlung by improving our modeling of the $\pi+\pi\to
\pi+\pi+\gamma$ process.
We will suggest a generalization to the other meson
species and  meson+baryon interactions in the end of this
section.

In order to calculate the differential cross sections for the photon
production in the processes of the type $\pi+\pi\to \pi+\pi+\gamma$,
we use the one-boson exchange (OBE) model as  originally applied
in Ref.~\cite{Eggers:1995jq} to the dilepton bremsstrahlung in pion+pion
collisions, later on in Ref.~\cite{Liu:2007zzw} to the low-energy
photon bremsstrahlung in pion+pion and kaon+kaon collisions.

In order to achieve a reliable level of accuracy we model the
interactions of pions with hadrons using a covariant microscopic
effective theory with the interaction Lagrangian,
\be L_{int}= g_{\sigma} \sigma \partial _\mu \vec \pi \partial ^\mu
\vec \pi + g _\rho \vec \rho ^\mu \cdot (\vec \pi \times
\partial _\mu \vec \pi) + g_f f_{\mu \nu} \partial ^\mu \vec \pi
\cdot
\partial ^\nu \vec \pi, \label{Lagrangian} \ee
as suggested in Refs.~\cite{Eggers:1995jq,Haglin:1992fy}.
Within this model the interaction of pions is described by the
exchange of scalar, vector and tensor resonances: $\sigma$, $\rho$
and $f_2(1270)$, respectively. Additionally, the form factors are
incorporated in the vertices in the t- and u-channels to account for
the composite structure of the mesons and thus to
effectively suppress the high momentum transfers:
\be h_{\alpha}(k^2)=\frac{m_\alpha^2-m_\pi^2}{m_\alpha^2-k^2}, \ee
where $m_\alpha=m_\sigma$ or $m_\rho$ or $m_f$ is the mass of the
exchanged meson and $k^2$ is the momentum transfer squared.


The cross section for $\pi+\pi\to\pi+\pi$ scattering is given by
\be \frac{d \sigma_{el} (s)}{dt} = \frac{|M_{el}|^2}{16 \pi s
(s-4m_\pi^2)}, \label{el1} \ee
where the matrix element $|M|^2$ is calculated by coherently
summing up
the Born diagrams of the $\sigma$-, $\rho$- and $f_2$-meson exchange
in $t$, $s$ and $u$  channels
(the $u$-channel diagrams are needed only in case of identical
pions), 
\bea |M_{el}|^2&=& |M^s(\sigma)\!+\! M^t(\sigma)\! +\! M^u(\sigma) \nonumber \\
&& +\!M^s(\rho)\!+\!M^t(\rho)\!+\!M^u(\rho)\!
\nonumber \\
&&  +\!M^s(f)\!+\!M^t(f)\! +\!M^u(f)|^2  . \ \ \ \ \ \ \ \
 \eea
Let us define the four-momenta of the incoming pions as
$p_a=(E_a,\vec p_a)$ and $p_b=(E_b,\vec p_b)$, the momenta of the
outgoing pions as $p_1=(E_1,\vec p_1)$ and $p_2=(E_2,\vec p_2)$ and
the four-momentum of the exchanged resonance ($\sigma$, $\rho$ or
$f_2$) as $k$.

The propagators of the massive and broad scalar and vector particles
are used to describe the exchange of the $\sigma$ and $\rho$ mesons
(see e.g. Ref.~\cite{Eggers:1995jq}). The resonance $f_2$ is a
spin-2 particle, for which the full momentum-dependent propagator
has been derived in~\cite{vanDam:1970vg}. The polarization sum is
\bea P_{\mu \nu \alpha\beta}  & = & \frac{1}{2} (g_{\mu\alpha}
g_{\nu\beta} + g_{\mu\beta} g_{\nu\alpha} - g_{\mu\nu}
g_{\alpha\beta}  ) \nn  && \!\!\!\!\!\!\!\!\! - \frac{1}{2}
(g_{\mu\alpha}\frac{k_\nu k_\beta}{m_f^2} +  g_{\mu\beta}\frac{k_\nu
k_\alpha}{m_f^2} + g_{\nu\alpha}\frac{k_\mu k_\beta}{m_f^2} +
g_{\nu\beta}\frac{k_\mu k_\alpha}{m_f^2}  ) \nn &&
\!\!\!\!\!\!\!\!\! + \frac{2}{3} (\frac{1}{2} g_{\mu \nu} +
\frac{k_\mu k_\nu}{m_f^2} )(\frac{1}{2} g_{\alpha \beta} +
\frac{k_\alpha k_\beta}{m_f^2} ). \eea
Following the example of the dilepton production study in
Ref.~\cite{Eggers:1995jq}, we use the same propagator for the $f_2$
resonance while additionally accounting for its finite width by
adding an imaginary part to the self-energy in accordance with the lifetime.

\begin{figure}
\includegraphics[width=0.45\textwidth]{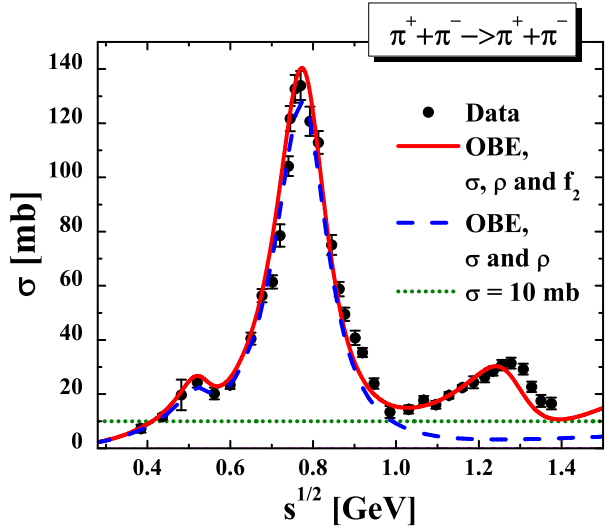}
\caption{(Color on-line) Cross section of pion+pion elastic
scattering within two effective models are compared to the
experimental data
from~\protect{\cite{Srinivasan:1975tj,Protopopescu:1973sh}}: the
exchange of  two mesonic resonances, scalar $\sigma$ and vector
$\rho$ (blue dashed line), and the exchange of  three resonances
$\sigma$, $\rho$ and the tensor resonance $f_2(1270)$ of the
particle data booklet~\cite{PDG} (red solid line). The green dotted
line shows the constant $\sigma_{el}=10$~mb for comparison.
\label{pipi}}
\end{figure}

\begin{figure*}
\includegraphics[width=0.9\textwidth]{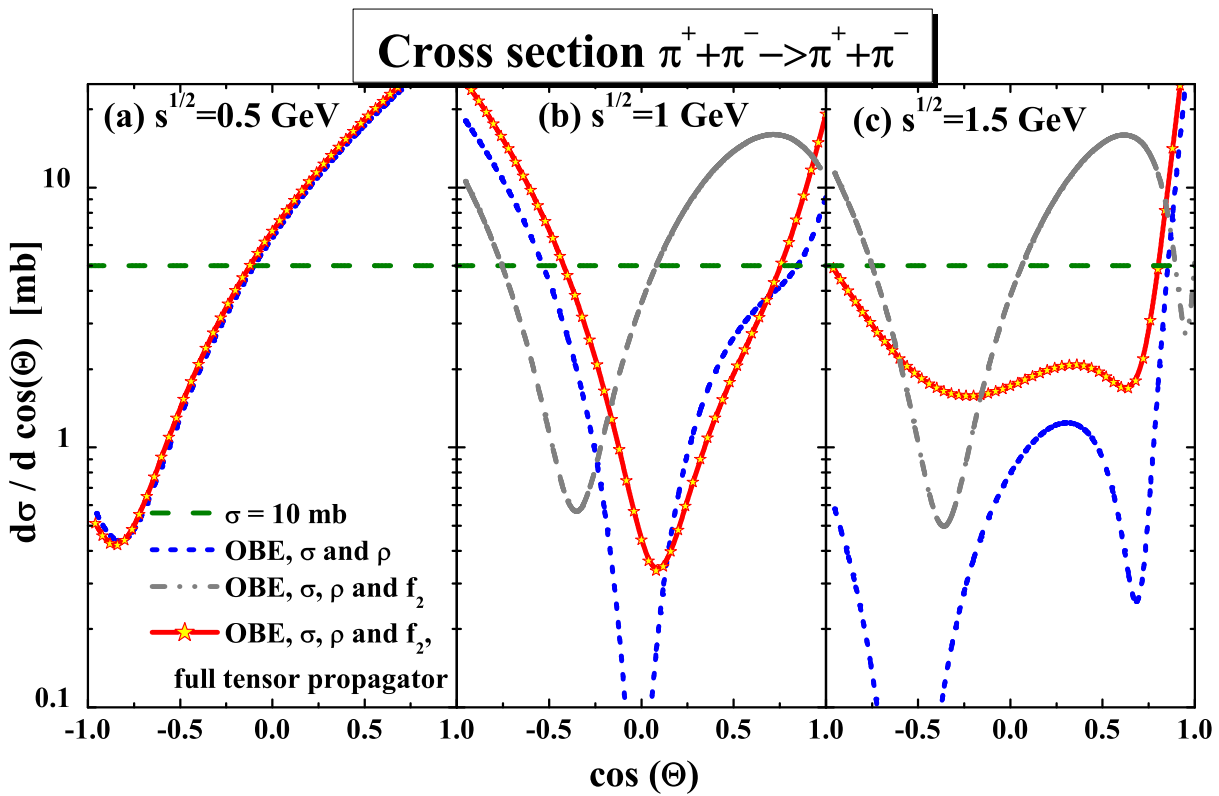}
\caption{(Color on-line) Angular differential cross section for
pion+pion elastic scattering within two effective models: the
exchange of  two mesonic resonances, scalar $\sigma$ and vector
$\rho$ (blue dashed line), and the exchange of  three resonances
$\sigma$, $\rho$ and the tensor particle $f_2(1270)$ (red solid line
with symbols and gray dashed line). The red solid line with star
symbols shows the model with the full momentum dependence of the
$f_2$ propagator, while the grey dashed line is obtained neglecting
the momentum dependence of the $f_2$ propagator. The green dashed
line shows the constant and isotropic $\sigma_{el}=10$~mb for
comparison. \label{angular}}
\end{figure*}

As the result, the following expressions are obtained for the matrix
elements of the elastic $\pi+\pi$ scattering diagrams (we give here
explicitly the $t$- and $s$-channel results, the $u$-channels can be
easily obtained by the crossing relations):
\bea M^t(\sigma) & = &  \frac{-g_\sigma^2 h_\sigma^2(t) \left( 2
m_\pi^2-t \right)^2}{t-m_\sigma^2+im_\sigma \Gamma_\sigma}, \nn
M^s(\sigma) & = & \frac{-g_\sigma^2 \left(s - 2 m_\pi^2
\right)^2}{s-m_\sigma^2+im_\sigma \Gamma_\sigma}, \nn M^t(\rho) & =
& \frac{-g_\rho^2 h_\rho^2(t) \left(
s-u\right)^2}{t-m_\rho^2+im_\rho \Gamma_\rho}, \nn M^s(\rho) & = &
\frac{g_\rho^2 \left( u-t\right)^2}{s-m_\rho^2+im_\rho \Gamma_\rho},
\label{elasticformulas} \\ M^t(f) & = & \frac{g_f^2 h_f^2(t)
}{t-m_f^2+im_f \Gamma_f} \nn &&
\!\!\!\!\!\!\!\!\!\!\!\!\!\!\!\!\!\!\!\! \times \frac{1}{2} \left(
\frac{2}{3} (2 m_\pi^2-t)^2-(s-2 m_\pi^2)^2 - (2m\pi^2-u)^2 \right),
\nonumber \eea \bea  M^s(f) & =  & \frac{g_f^2 }{s-m_f^2+im_f
\Gamma_f} \nn && \!\!\!\!\!\!\!\!\!\!\!\!\!\!\!\!\!\!\!\! \times
\frac{1}{2} \left( \frac{2}{3} (s-2 m_\pi^2)^2-(2 m_\pi^2-t)^2 -
(2m\pi^2-u)^2 \right), \nonumber
\eea
where the Mandelstamm variables are defined as
$s=(p_a+p_b)^2=(p_1+p_b)^2$, $t=(p_a-p_1)^2=(p_b-p_2)^2$,
$u=(p_a-p_2)^2=(p_b-p_1)^2$.

We point out that the formulae (\ref{elasticformulas}) are compact,
because the masses of all pions were assumed to be equal to $m_\pi$
and the energy-momentum conservation $p_a+p_b=p_1+p_2$ was assumed.
These conditions are not satisfied for the off-shell
$\pi+\pi\to\pi+\pi$ subprocess,  which  we will encounter in the
subsequent calculation of the bremsstrahlung photon production
$\pi+\pi\to\pi+\pi+\gamma$. For the actual calculation we will use
the off-shell generalizations $M(p_a,p_b,p_1,p_2)$ of the formulae
(\ref{elasticformulas}), which we derive from the Feynman diagrams,
but the off-shell formulae are too lengthy to be presented here.

A reduced version of the model with the exchange of only two
resonances -- the scalar $\sigma$ and the vector $\rho$ meson -- was
used by the authors of Ref.~\cite{Liu:2007zzw} to calculate the rate
of the photon production from the $\pi+\pi\to\pi+\pi+\gamma$ process
at low transverse momenta of the photons ($p_T<0.4$~GeV). This
approximation is suitable at low $p_T$ because the photon rate in
this kinematical region is dominated by  pion collisions of low
center-of-mass energy $\sqrt{s}$, for which the contribution of the
$f_2$-exchange is small. However, relatively high transverse momenta
of photons $p_T=1-2$~GeV are of interest for our goal of clarifying
the "puzzling" high elliptic flow of direct photons. Thus we need a
robust model for the interaction of mesons  also at
$\sqrt{s}>1$~GeV. Therefore, we use the OBE model with three mesons
as interaction carriers (including the tensor particle $f_2(1270)$)
in our present calculations.

Phenomenological coupling constants, masses and widths of the three
interaction-carriers that enter the Lagrangian (\ref{Lagrangian})
have to be fixed to the integrated energy-dependent cross section of
the pion+pion elastic scattering $\sigma_{el} (s)$, which is known
experimentally. We present in Fig.~\ref{pipi} the integrated cross
section of the $\pi+\pi$ elastic scattering in the two versions of
the OBE model described above: taking into account  2 resonances
$\sigma$, $\rho$ (dashed blue line) and taking into account 3
resonances $\sigma$, $\rho$ and $f_2$ (solid red line). Fitting the
parameters of both variants of the OBE model (with two- or
three-resonance exchange), the data
from~\cite{Srinivasan:1975tj,Protopopescu:1973sh} can be described.
The best-fit parameters for the model used here are: $g_\sigma
m_\sigma = 2.0$, $m_\sigma= 0.525$~GeV, $\Gamma_\sigma=0.100$~GeV,
$g_\rho=6.15$, $m_\rho=0.775$~GeV, $\Gamma_\rho=0.15$~GeV, $g_f
m_f=8.0$, $m_f=1.274$~GeV, $\Gamma_f=0.18$~GeV. 
The values of the masses and widths suggest the identification of
the $\rho$-resonance to the $\rho$-meson and of the particle $f_2$
to the $f_2(1270)$ in the particle data book~\cite{PDG}.

One sees in Fig.~\ref{pipi} that the tensor particle $f_2$ is
important for the description of the pion interaction at higher
collision energies $\sqrt{s}>1$~GeV. Neglecting the contribution of
the $f_2$ leads to an underestimation of the $\pi+\pi$ elastic
scattering cross section by an order of magnitude around
$\sqrt{s}=1.2-1.3$~GeV. Later data on the $\pi+\pi$ interaction at
$\sqrt{s}$ above 1~GeV -- extracted in Ref.~\cite{Aston:1990wg} from
the measurement of the $K+p\to\Lambda+\pi+\pi$ reaction -- also
point to the importance of the tensor interaction in the resonance
region of the $f_2(1270)$.
%

With the parameters fixed to the integrated pion elastic scattering
cross section as described above, we can calculate the {\em
differential} cross section $d\sigma_{el}(s)/dt$, which is a
necessary ingredient for the calculation of the bremsstrahlung
photon production within the soft-photon
approximation~\cite{Haglin:1992fy} for the low photon energy
and low $\sqrt{s}$ (see below). The results of our calculations for
the differential cross section of the process $\pi+\pi\to\pi+\pi$ as
a function of the scattering angle $\Theta$ in the $\pi+\pi$
center-of-mass system is presented in Fig.~\ref{angular} within two
effective OBE models: including the exchange of  two mesonic
resonances, scalar $\sigma$ and vector $\rho$ (blue dashed line), and
including the exchange of  three resonances $\sigma$, $\rho$ and the
tensor particle $f_2$ (red solid line with symbols and gray dashed
lines). The red solid line with star symbols shows the latter model
with the full momentum dependence of the $f_2$ propagator, while the
grey dashed line is obtained, if the momentum dependence of the $f_2$
propagator is neglected. The green dashed line shows the constant
and isotropic $\sigma_{el}=10$~mb for comparison.

\begin{figure*}
\includegraphics[width=0.9\textwidth]{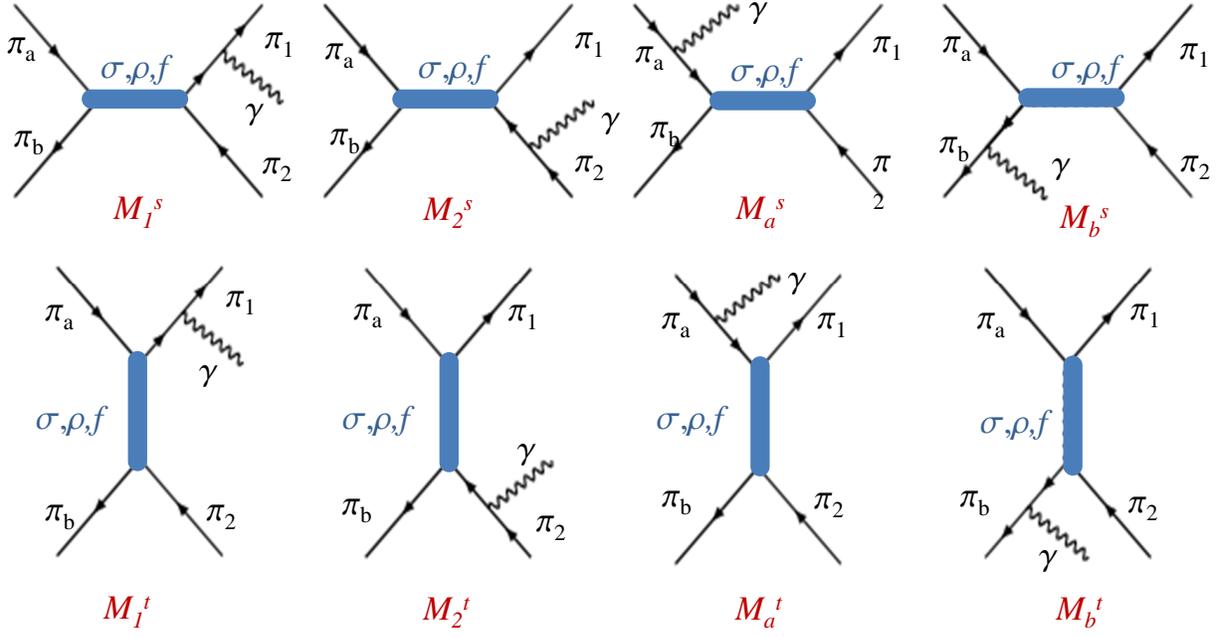}
\caption{(Color on-line) Feynman diagrams for photon production in
the reaction $\pi+\pi\to\pi+\pi+\gamma$ in the one-boson exchange
(OBE) model. The time goes from left to right. For  identical pions,
e.g. $\pi^+ + \pi^+$, the $u$-channel diagrams have to be added.
\label{diagramsOBE} }
\end{figure*}

One can see the pronounced influence of the tensor resonance
exchange at higher $\sqrt{s}$ of the collision. One observes
that for low $\sqrt{s}=0.5$~GeV the $\sigma$ peak dominates the
cross section and the angular dependence is governed fully by the
combination of scalar and vector terms. On the other hand, the
influence of the $f_2$-resonance is clearly seen at $\sqrt{s}=1$~GeV
and $\sqrt{s}=1.5$~GeV: the model with two resonances (dashed blue
line) differs from the results of the other OBE variants. The
correct momentum dependence of the $f_2$ propagator is important, as
one sees from the difference of the star red and grey dashed lines.

Figs. \ref{pipi} and~\ref{angular} demonstrate that the inclusion of the
tensor resonance in the model leads not only to an increase of the
cross section for $\sqrt{s}>1$~GeV (in agreement with the data) but
also to a considerable "flattening" of the $\cos{\Theta}$
distribution. The cross section at higher $\sqrt{s}$ in the extended
model is more isotropic.
%

Using the  OBE model (described above)  for the covariant
interactions of pions, we can calculate the emission of photons by
the colliding pions by gauge coupling to the external hadron lines.
The Feynman diagrams for the photon production in the process
$\pi+\pi\to\pi+\pi+\gamma$ are shown in Fig.~\ref{diagramsOBE}. For
identical pions, e.g. $\pi^+ + \pi^+$, the $u$-channel diagrams have
to be added, which are obtained from the $t$-channel diagrams by
exchanging the outgoing pions. The applicability of this method is
not limited by the low energy of the photon but is restricted only
by the applicability of the effective model to the description of
the pion (elastic) interaction.

Let us again denote the four-momenta of the incoming pions by $p_a$
and $p_b$, the momenta of the outgoing pions by $p_1$ and $p_2$,
while the photon momentum will be denoted by $q=(q_0,\vec q)$. The
cross section for photon production in the process
\be \pi(p_a) + \pi(p_b) \to \pi(p_1) + \pi(p_2) +\gamma(q) \ee
is given by
\be d \sigma^{\gamma} = \frac{1}{2\sqrt{s(s-4m_\pi^2)}}
|M(\gamma)|^2 d R _3, \label{r3} \ee
where $d R_3$ is the three-particle phase space, which depends on
the momenta of the outgoing pions and of the photon,
\bea dR_3 & = & \frac{d^3p_1}{(2\pi)^3 2 E_1} \frac{d^3p_2}{(2\pi)^3
2 E_2} \frac{d^3q}{(2\pi)^3 2 q_0} \nn &&
 (2\pi)^4 \delta^4\left(
p_a+p_b-p_1-p_2-q\right).
 \eea
The cross section (\ref{r3}) will be integrated over the final pion
momenta to obtain $d\sigma/d^3q$. The $\delta$-function allows to perform four
integrations analytically and the remaining two are done numerically.

The matrix element $M$ in (\ref{r3}) is a coherent sum of the
diagrams presented in Fig.~\ref{diagramsOBE} -- i.e. of the photon
attached to each pion line $\pi_a$, $\pi_b$, $\pi_1$ and $\pi_2$ --
and of contact terms, which account for the emission from the
vertices and the internal lines:
\bea |M(\gamma)|^2 &=& M^*_\mu(\gamma) M^\mu(\gamma) \nn &=& \left|
M_a^\mu+M_b^\mu + M_1^\mu+M_2^\mu + M_c^\mu \right|^2.
 \eea
The complex matrix elements for the photon emission from each of the
pion lines $M_i^\mu$ are calculated as sums of the three meson
exchanges ($\sigma$, $\rho$, $f_2$). For instance:
\bea M^\mu _1 &=& e J_1^\mu \left[ M^s_{el}(p_a,p_b,p_1+q,p_2)
\right. \nn && + M^t_{el}(p_a,p_b,p_1+q,p_2) \nn && \left. +
M^u_{el}(p_a,p_b,p_1+q,p_2) \right], \eea
with
\bea J_{a,b}^\mu= -Q_{a,b}\frac{(2 p_{a,b}-q)^\mu}{2 p_{a,b}\cdot
q}, \nn J_{1,2}^\mu = Q_{1,2}\frac{(2 p_{1,2}-q)^\mu}{2 p_{1,2}\cdot
q}, \label{charges2} \eea
where $Q_i$ are the charges of the pions in terms of the electron
charge $e$. The matrix elements for the pion elastic subprocess
$M_{el}(p_a,p_b,p_1+q,p_2)$ are the off-shell generalizations of the
formulae (\ref{elasticformulas}).

The contact term $M_c^\mu$ is taken from Ref.~\cite{Liu:2007zzw},
eq. (14), where it was derived by demanding the gauge invariance of
the resulting cross section. Indeed, the gauge invariance of the
result often has to be restored~\cite{Haglin:1989ga} in calculations
within effective models. In the present work, we have used the
contact terms in order to cancel the gauge-dependent parts in the
matrix element as
 in  Ref.~\cite{Liu:2007zzw}. Alternatively, one can take
into account additional diagrams with the emission of photons from
the internal lines (see Refs~\cite{Eggers:1995jq}) but this method
does not always eliminate the need for contact terms (see
Ref.~\cite{Haglin:1989ga}). We have verified that $q_\mu M^\mu
(\gamma) = 0$.
Comparing our results to calculations with a different gauge-fixing
method will allow to quantify the uncertainty of the effective model
applied (work in progress).

\begin{figure}
\includegraphics[width=0.45\textwidth]{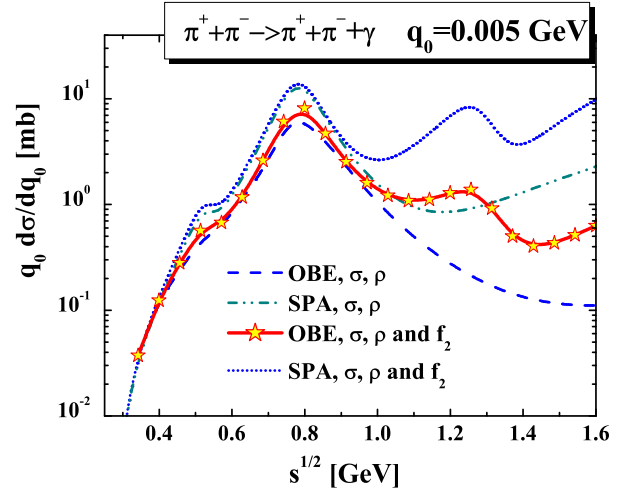}
\caption{(Color on-line) Cross section for the production of a
photon with energy $q_0=0.005$~GeV in the process
$\pi+\pi->\pi+\pi+\gamma$ within the following models: the exact OBE
cross section within the effective model taking into account scalar,
vector and tensor interactions via the exchange of $\sigma$, $\rho$
and $f_2(1270)$-mesons gives (red line with star symbols), the soft
photon approximation to this model (blue dotted line); the OBE result
within the model taking into account only the scalar and vector
interactions via the exchange of $\sigma$ and $\rho$ mesons (blue
dashed line), and the soft photon approximation to this model (cyan
dash-dot-dotted line). \label{XsectE0005}}
\end{figure}


Within the soft-photon approximation (SPA)~\cite{Gale87} one assumes
that the strong interaction vertex is on-shell. The soft photon
approximation is based on the first-order expansion in the Low
theorem~\cite{Low:1958sn} and is valid at low photon energy and low
$\sqrt{s_{mm}}$ of the meson+meson collision, as has been studied in
detail for the production of dileptons in Ref.~\cite{Eggers:1995jq}.
In this case the strong interaction part and the electromagnetic
part can be separated, i.e. the soft-photon cross section for the
reaction $m_1 + m_2 \to m_1 + m_2 + \gamma$ can be written as
\be q_0 \frac{d\sigma^{\gamma}(s)}{d^3q} = \frac{\alpha_{_{EM}}}{4
\pi^2} \!\!\!\! \int\limits_{-\lambda(s,m_a^2,m_b^2)/s}^0
\!\!\!\!\!\!\!\!\!\! |\epsilon \cdot J(q,t)|^2 \frac{d\sigma_{el}
(s)}{dt} dt, \label{int_spa_formula} \ee
where $\alpha_{_{EM}}$ is the fine structure constant, $t$ is the
momentum transfer squared  in the $\pi+\pi\to\pi+\pi$ sub-process,
and $\epsilon$ is the photon polarization. $J^\mu$ is the
electro-magnetic current
\be J^\mu=-Q_a\frac{p_a^\mu}{(p_a\cdot q)}
-Q_b\frac{p_b^\mu}{(p_b\cdot q)} +Q_1\frac{p_1^\mu}{(p_1\cdot q)}
+Q_2\frac{p_2^\mu}{(p_2\cdot q)}. \nonumber \ee
The polarization sum
\be |\epsilon \cdot J|^2 = \left\{ \sum_{pol \ \lambda } J\cdot
\epsilon_\lambda  J\cdot \epsilon_\lambda \right\} \ee
depends on the photon momentum $q$, the charges of the pions $Q_i$
as well as on the invariant kinematic variables, including $t$. For
the case of equal-mass particle scattering
($m_a=m_b=m_1=m_2=m_\pi$), one obtains~\cite{Haglin:1992fy}:
\bea |\epsilon\cdot J|^2 & = & \frac{1}{q_0^2}\left\{
-(Q_a^2+Q_b^2+Q_1^2+Q_2^2) \right. \nn
&&  \hspace{-1.8cm} \left.
-2(Q_aQ_b\!+\!Q_1Q_2)\frac{s-2m_\pi^2}{\sqrt{s(s-4m_\pi^2)}} \ln
\left(\frac{\sqrt{s}+\sqrt{s-4m_\pi^2}}{\sqrt{s}-\sqrt{s-4m_\pi^2}}
\right)
\right. \nn
&&  \hspace{-1.8cm} \left.
+2(Q_aQ_1\!+\!Q_bQ_2)\frac{2m_\pi^2-t}{\sqrt{t(t-4m_\pi^2)}} \ln
\left(\!\!
\frac{\sqrt{-t+4m_\pi^2}+\sqrt{-t}}{\sqrt{-t+4m_\pi^2}-\sqrt{-t}}
\right)
 \right. \nn
&&  \hspace{-1.8cm} \left.
+2(Q_aQ_2\!+\!Q_bQ_3)\frac{s-2m_\pi^2+t}{\sqrt{(s+t)(s+t-4m_\pi^2)}}
\right. \nn
&&  \hspace{1.cm} \left. \times \ln
\left(\frac{\sqrt{s+t}+\sqrt{s+t-4m_\pi^2}}{\sqrt{s+t}-\sqrt{s+t-4m_\pi^2}}
\right) \right\}. \label{charges1} \eea
In (\ref{int_spa_formula}), $d \sigma_{el}(s)/dt$ is the on-shell
differential elastic $\pi+\pi$ cross section, which is a function of
the invariant energy $s$ and the pion scattering angle via $t$.

The expression (\ref{int_spa_formula}) is considerably simpler in
comparison to the ``exact" OBE formula (\ref{r3}) because of the
factorization of the diagrams from Fig.~\ref{diagramsOBE} into the
electromagnetic part and the elastic $\pi+\pi\to\pi+\pi$
sub-process, for the cross section of which the $q$-dependence is
omitted. This corresponds to neglecting the off-shellness of the
pion, which emits the photon, e.g. for the pion $a$:
\be p_a-q\approx p_a. \ee
Consequently, the sub-process invariant energy $s_2$ is also
approximated by the total invariant energy of the process
$\pi+\pi\to\pi+\pi+\gamma$:
\be s_2\equiv (p_a+p_b-q)^2 \approx (p_a+p_b)^2 = s, \ee
and the limits of integration over $t$ are also taken as for the
on-shell case, i.e. from $-\lambda(s,m_a^2,m_b^2)/s$ to 0, while the
actual integration over the full 3-particle phase space in the exact
treatment (\ref{r3}) involves different limits for $t$.

In Fig.~\ref{XsectE0005}
we show the resulting cross sections for the photon production in
the process $\pi+\pi\to\pi+\pi+\gamma$ within the following models:
the ``exact" OBE taking into account scalar, vector and tensor
interactions via the exchange of $\sigma$, $\rho$ and
$f_2(1270)$-mesons gives the red line with star symbols, the soft
photon approximation (\ref{int_spa_formula}) to this model is shown
by the blue dotted line; the OBE result within the model taking into
account only the scalar and vector interactions via the exchange of
$\sigma$ and $\rho$ mesons is shown by the blue dashed line, and the
soft photon approximation to this model is presented by the cyan
dash-dot-dotted line. The photon energy is fixed to $q_0=5$~MeV.
%
For the very low energy of the photon of $q_0=5$~MeV the SPA agrees
with the ``exact" cross section very well in the region of
$\sqrt{s}<0.9$~GeV (see Fig.~\ref{XsectE0005}). However, the
discrepancy to the OBE result is increasing rapidly with growing
$\sqrt{s}$;  the calculations for the higher photon energy of
$q_0=0.5$~GeV show an even larger discrepancy between the SPA and
the exact OBE result.

\begin{figure}
\includegraphics[width=0.45\textwidth]{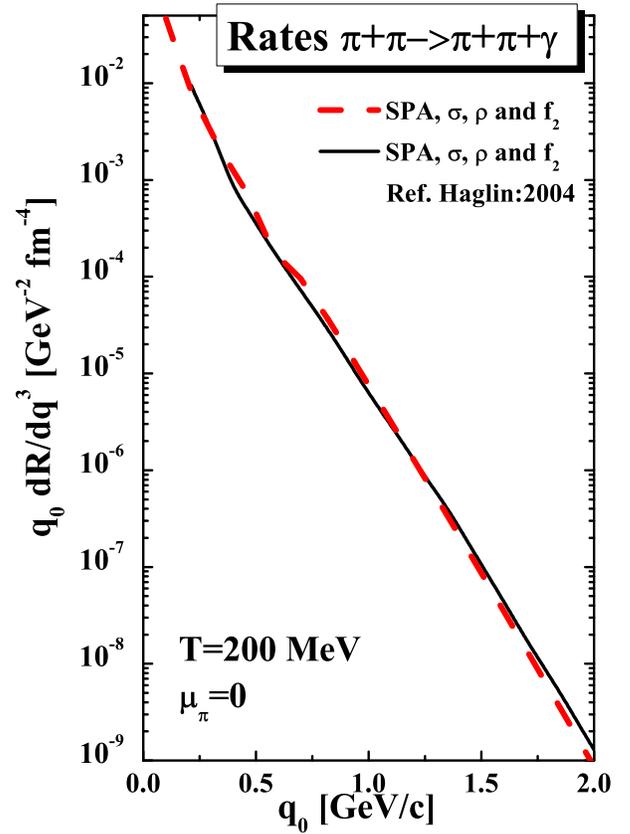}
\caption{(Color on-line) Invariant rate of the bremsstrahlung photon
production from an equilibrated pion gas at a temperature of
$T=200$~MeV and pion chemical potential $\mu_\pi=0$ as calculated in
the OBE model with three resonance exchange within the soft-photon
approximation (red dashed line). The black solid line "Haglin:2004"
from Ref.~\protect{\cite{Haglin:2003sh}} is shown for comparison.
\label{VariantsSPA} }
\end{figure}

Using the cross section for the $\pi+\pi\to\pi+\pi+\gamma$ reaction according
to (\ref{int_spa_formula}) as a function of the photon energy $q_0$
and the collision energy $\sqrt{s}$, we calculate the yield
$dN/d^3q$ and the invariant rate $q_0 dR/d^3q$ of bremsstrahlung
photon production from an equilibrated pion gas.
Within kinetic theory, the rate of photon production in the
collisions of particles $a$ and $b$ in a thermalized medium (number
of photons produced per unit space-time volume $d^4x$) is an
integral over the three-momenta of the incoming particles:
\bea q_0 \frac{dN}{dx^4 d^3q}  = g \int \! ds \int \! \frac{d^3
p_a}{(2\pi)^3} \int \! \frac{d^3 p_b}{(2\pi)^3} \, e^{-(E_a+E_b)/T}
\nn \times v_{rel} q_0 \frac{d\sigma^\gamma}{d^3q}
\delta(s-(p_a+p_b)^2), \ \ \label{kinetic} \eea
where $T$ is the temperature, $v_{rel}$ is the relative velocity
given by
\be v_{rel}=\frac{\sqrt{(p_a\cdot p_b)^2-m_a^2 m_b^2}}{E_a E_b} ,\ee
and $g=(2s_a+1)(2s_b+1)$ is the spin degeneracy factor. Integrating
the expression (\ref{kinetic}) over the particle momenta one
obtains~\cite{Haglin:1992fy}:
\be q_0  \frac{dN}{d^4xd^3q}= \frac{T^6 g}{16 \pi^4} \int
\limits_{z_{min}}^\infty \!\! dz
\frac{\lambda(z^2T^2,m_a^2,m_b^2)}{T^4} K_1(z)
q_0\frac{d\sigma^\gamma}{d^3q} , \ee
where $ z_{min} = (m_a+m_b)/T $, $z=\sqrt{s}/T$, and $K_1(z)$ is the
modified Bessel function.

The expression (\ref{kinetic}) can be generalized to account for
quantum effects such as Bose enhancement or Pauli blocking
(depending on the particle type) by integrating additionally over
the momenta of the final particles and changing the Boltzmann
distributions to the Fermi or Bose distribution functions $f_i(T)$:
\bea q_0 \frac{dN}{dx^4 d^3q}  = g \! \int \! ds \! \int \!
\frac{d^3 p_a}{(2\pi)^3} \! \int \! \frac{d^3 p_b}{(2\pi)^3} \! \int
\! \frac{d^3 p_1}{(2\pi)^3} \! \int \! \frac{d^3 p_1}{(2\pi)^3}
\nn f_a(T) f_b(T) (1-f_1(T)) (1-f_2(T))
\nn \times   v_{rel} \, q_0 \frac{d\sigma^\gamma}{d^3q} \,
\delta(s-\!(p_a+p_b)^2). \ \  \label{kinetic2} \eea
In the current section we calculate the thermal rates according to
formula (\ref{kinetic}). However, within the PHSD transport approach
for the heavy-ion collisions in section~\ref{sect:results}  the
effects of the quantum statistics will be taken into account.

In Fig.~\ref{VariantsSPA} the rates are presented for a temperature
$T=200$~MeV and pion chemical potential $\mu_\pi=0$ for the  OBE
model with three resonance exchanges adopting the soft-photon
approximation (red dashed line). We confirm the results from
Ref.~\cite{Haglin:2003sh} (black solid line) calculated within the
same assumptions (SPA, three resonances) but with a slightly
different  parameter set of the Lagrangian. It is, however,
questionable that the SPA is applicable at high photon energies.

We note that the accuracy of the SPA approximation can be
significantly improved and the region of its applicability can be
extended by slightly modifying the formula (\ref{int_spa_formula})
-- i.e. by evaluating the on-shell elastic cross section at the
invariant energy $s_2$ of the sub-process. The latter is
kinematically fixed to \be s_2=s-q_0 \sqrt{s} \ne s,\ee
Thus the modified SPA formula is
\be q_0 \frac{d\sigma^{\gamma}(s)}{d^3q} = \frac{\alpha_{_{EM}}}{4
\pi} \!\!\!\!\! \int\limits_{-\lambda(s_2,m_a^2,m_b^2)/s_2}^0 \!\!
\!\!\!\!\!\!\!\!\!\! |\epsilon \cdot J(q,t)|^2 \frac{d\sigma_{el}
(s_2)}{dt} dt. \label{int_spa_formula2} \ee
In the following, we will denote the approximation (\ref{int_spa_formula2})
as ``improved SPA" and will show below that it provides a good
description of the exact photon production rates.

But first we describe a simple model for the photon production in
meson+meson collisions by the bremsstrahlung mechanism, i.e. the
approximation of a constant $meson+meson$ elastic cross section. We
have seen above that the elastic-scattering cross section of pions
is approaching about $\sigma=10$~mb at high $\sqrt{s}$ and becomes
increasingly isotropic. As a very simple estimate for the
interaction of two mesons one can use a constant isotropic
elastic-scattering cross section of, e.g., $\sigma=10$~mb at all
$\sqrt{s}$. In case of the isotropic cross section $\sigma_{el}$, it
can be taken out of the integral (\ref{int_spa_formula}). The
integration of the electromagnetic current over the photon angle in
(\ref{int_spa_formula})  can be done in a straightforward way in
case of small momentum transfer~\cite{Haglin:1992fy}, since
\be \int_{tmin}^0 t \  dt = \frac{(s-4 m_{\pi}^2)^2 }{ 2} .\ee
This leads to a useful approximation to the cross section for the
photon bremsstrahlung  in meson+meson collisions
\begin{eqnarray}
q_0\frac{d^3 \sigma^\gamma}{d^3 q} & = & \frac{\alpha}{4 \pi}
\frac{{\bar \sigma (s)}}{q_0^2},
            \label{brems} \\
{\bar \sigma(s)} & = & \frac{s - (M_1 + M_2)^2}{2 M_1^2} \sigma(s),
\label{brems2}
\end{eqnarray}
where $M_1$ and $M_2$ are the masses of the colliding mesons; here
the first meson is assumed to have unit charge and the second meson
to be charge-neutral, i.e. the approximation does not take into
account the full dependence on the charges of the pions as was done
in (\ref{charges1}) or (\ref{charges2}). This
``constant-cross-section" approximation is useful in particular for
an estimate of the photon bremsstrahlung in elastic collisions of
mesons, for which the scattering cross sections are not known
experimentally. We recall that the formulae (\ref{brems}) and
(\ref{brems2}) have been used to model photon bremsstrahlung in the
collisions of various meson species in the previous transport
calculations in
Refs.~\cite{Bratkovskaya:2008iq,Linnyk:2013hta,Linnyk:2013wma}.

\begin{figure}
\begin{center}
\includegraphics[width=0.45\textwidth]{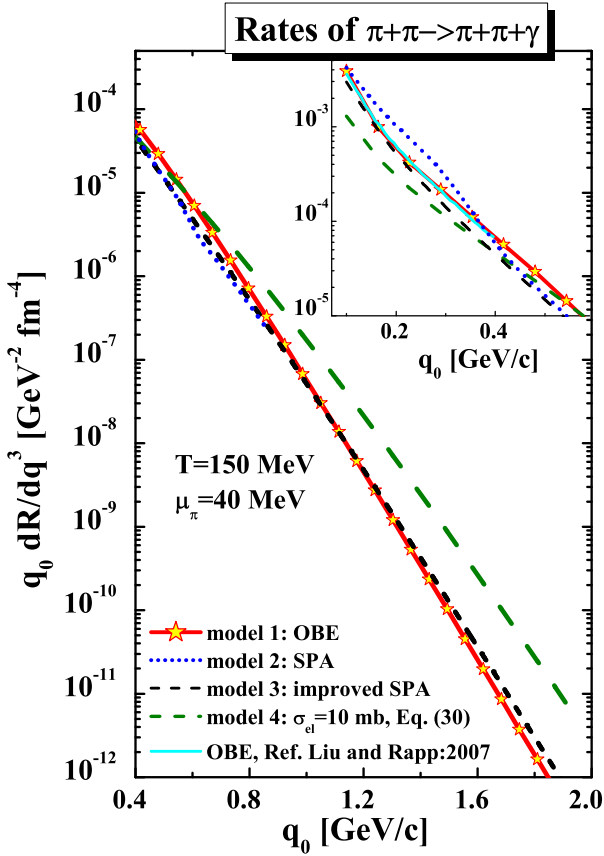}
\caption{(Color on-line) Invariant rate of bremsstrahlung photons
produced from an equilibrated pion gas at $T=150$~MeV and
$\mu_\pi=40$~MeV versus the photon energy $q_0$. The in-let shows
the same quantity for the range of photon energies
$q_0=0.1-0.4$~GeV. The calculations have been performed within the
following models: (1) OBE model beyond the soft-photon approximation
(red solid line with star symbols); (2) OBE model within the soft
photon approximation (blue dotted line); (3) OBE model within the
improved soft photon approximation (black short-dashed line) -- the
invariant energy $s_2$ of the on-shell $\pi+\pi$ elastic process is
not equal to the total invariant energy of the process $s$:
$s_2=s-q_0 \sqrt{s}$; (4) the soft photon approximation with the
constant isotropic elastic cross section of $\sigma_{el}=10$ using
the formula (\protect\ref{brems}) (green dashed line). The cyan
solid line "Liu and Rapp:2007" from
Ref.~\protect{\cite{Liu:2007zzw}} is shown for comparison.
\label{rates}}
\end{center}
\end{figure}

Finally, we proceed to calculate the photon production rates beyond
the soft photon approximation, using the cross section for the
$\pi+\pi\to\pi+\pi+\gamma$ reaction calculated according to the
exact OBE expression (\ref{r3}).
We present the calculated invariant rate $q_0 d R/dq^3$ of
bremsstrahlung photons produced from an equilibrated pion gas at
$T=150$~MeV and $\mu_\pi=40$~MeV in Fig.~\ref{rates}. The results of
the following models are compared:
\begin{itemize}
\item model 1 (red solid line): exact rates within the one-boson exchange model (OBE) {\em
beyond} the soft-photon approximation  -- i.e. using the formula
(\ref{r3}) for the photon production cross section $q_0\
d\sigma^\gamma/d^3q$;
\item
model 2 (blue dotted line): result within the soft photon approximation
-- i.e. using the formula (\ref{int_spa_formula}) -- while using the
elastic $\pi+\pi$ cross section calculated within the OBE model as
given by the equations (\ref{el1})-(\ref{elasticformulas});
\item
model 3 (black short-dashed line): results of the {\em improved} soft
photon approximation
 -- i.e. using the formula (\ref{int_spa_formula2})
instead of (\ref{int_spa_formula}) -- and the same pion elastic
scattering cross section as in the model 2;
\item
model 4 (green dashed line): soft photon approximation using a
constant isotropic elastic cross section of $\sigma_{el}=10$~mb and
assuming for the pion charges $Q_a=Q_1=1$, $Q_b=Q_2=0$ -- i.e. using
the formula (\ref{brems}). For this case the elastic cross section
does not depend on $\sqrt{s}$ and therefore there is no difference
between the SPA and improved SPA.
\end{itemize}
The rate of bremsstrahlung photons at low transverse momenta
$p_T<0.4$~GeV has been calculated before in
Ref.~\protect{\cite{Liu:2007zzw}} within the one-boson exchange
model with the exchange of two resonances for the same system. This
previous result is shown for comparison by the cyan dashed line and
is confirmed by our present calculations. The agreement is expected,
since our calculations differ only in the inclusion of the
$f_2$-meson exchange, which is important for large $\sqrt{s}$ and
does not play an important role for the production of low transverse
momentum photons, which is dominated by low $\sqrt{s}$ of the
$\pi+\pi$ collisions.

On the other hand, the SPA (model 2) deviates from the exact OBE
result (model 1) even at low $q_0$, because the former directly
follows the $\sqrt{s}$ structure of the elastic $\pi\pi$ cross
section. Since the formula (\ref{int_spa_formula}) does not account
for the off-shellness of the emitting pion, it overweights the
high-$\sqrt{s}$ part of the elastic cross section, in line with the
findings of Refs.~\cite{Haglin:1992fy,Eggers:1995jq}. We note that
the  OBE model presented here is constrained by the pion scattering data
only up to  $\sqrt{s_{\pi\pi}}=1.4$~GeV and generally cannot be
extended to large $\sqrt{s}$. Thus the SPA scenario "model 2" is not
reliable for large $q_0$ (approximately for $q_0>0.8$~GeV). This is
not the case for the improved SPA (model 3).

\begin{figure*}
\begin{center}
\includegraphics[width=\textwidth]{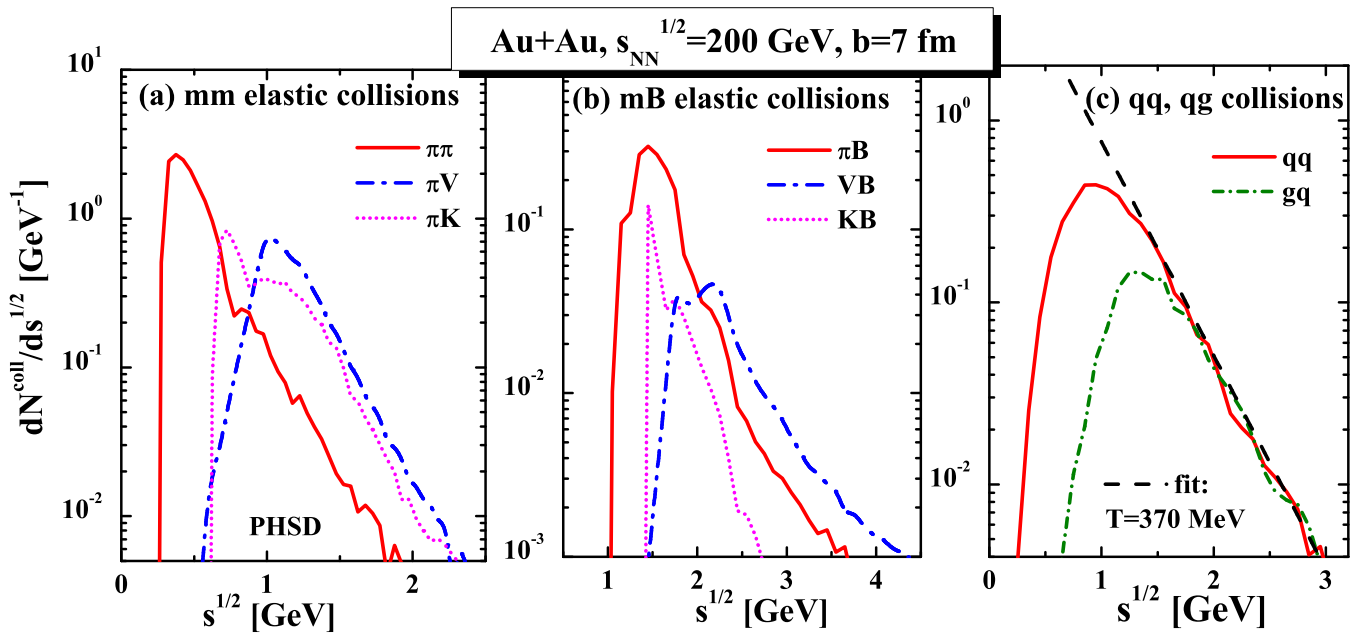}
\caption{(Color on-line) The distribution in the invariant collision
energy $\sqrt{s}$ for the elastic scattering of mesons on mesons
(left panel), mesons on baryons (middle panel) and for the elastic
scattering of partons (right panel) in the course of a $Au+Au$
collision at $\sqrt{s_{NN}}=200$~GeV for $b=7$~fm as calculated
within the PHSD approach. Here $V\equiv (\rho, \omega,
\phi)$; $K$ denotes all the strange mesons $K\equiv (K, \bar K, K^*,
\bar K^*)$ and $B$ stands for the baryons $B=(p, n, \Delta, \dots)$.
}\label{dNdS}
\end{center}
\end{figure*}

One can see in Fig.~\ref{rates} that the {\em improved} SPA
(\ref{int_spa_formula2}) gives a very good approximation to the
exact result at higher photon energies of up to  $q_0 \approx 2$~GeV. This
is because the $\sqrt{s_2}$ of the sup-process does not reach such
high values as $\sqrt{s}$, and the OBE model for the elastic cross
section is sufficiently realistic in this region of $\sqrt{s_2}$.

In comparison, the constant cross-section approximation (based on
formula (\ref{brems})) overestimates the exact rates for $q_0>1$~GeV
and underestimates for $q_0<0.4$~GeV. This model approximately
corresponds to the procedure used previously in our transport
calculations for the estimation of the photon bremsstrahlung in
meson+meson collisions in
Refs.~\cite{Bratkovskaya:2008iq,Linnyk:2013hta,Linnyk:2013wma}. In
the present work, we will use the exact OBE cross section $d
\sigma^\gamma/d^3q$. Therefore, we will now find a lower yield of
bremsstrahlung photons for $q_0>1$~GeV in the transport
simulations of heavy-ion collisions, see section~\ref{sect:results}
below.

Another reasons for the good agreement between the ``improved" SPA
rates with the exact ones is that the dominant contribution to the
rates comes from the low collision energies $\sqrt{s}$, while the
deviation between the exact cross section of the process
$\pi+\pi\to\pi+\pi+\gamma$ from that calculated within the improved
SPA is most pronounced at high $\sqrt{s}$ and high $q_0$. Such high
collision energies $\sqrt{s}$ are suppressed in a thermal medium
exponentially by the Boltzman factor describing the occupation of
the pion energies at fixed temperature.
It is not clear, whether this high accuracy of the ``improved SPA"
holds also out-of-equilibrium. We show in Fig.~\ref{dNdS} explicitly
the distributions in the number of meson+meson, meson+baryon and
parton+parton collisions versus their invariant energy $\sqrt{s}$ as
calculated within the PHSD for the case of a $Au+Au$ collision with
the energy $\sqrt{s_{NN}}=200$~GeV at impact parameter $b=7$~fm,
which is the centrality qualitatively similar to minimum bias. The
distribution of meson+meson, meson+baryon and parton+parton
collision invariant energies $dN/d\sqrt{s}$ in an actual heavy-ion
reaction shows a clear dominance of the low-$\sqrt{s}$ comover
collisions. Therefore we expect that the deviations of the cross
section $\pi+\pi\to\pi+\pi+\gamma$ in the improved SPA from the
exact ones at high $\sqrt{s}$ will not influence much the yield of
the photons that result from the integration over all pion
collisions:
\be q_0 \frac{d^3 \sigma^\gamma(AA)}{dq^3} = \int \! ds \,
\frac{dN^{coll}_{\pi\pi}}{ds} \, q_0 \frac{d ^3\sigma^\gamma
(\pi\pi)}{dq^3}. \ee

In summarizing the results of the current section, we have improved
the implementation of the photon bremsstrahlung in the process
$\pi+\pi\to\pi+\pi+\gamma$ within the PHSD transport approach, using
now the exact OBE cross sections beyond the soft-photon
approximation. The bremsstrahlung photon production in collisions of
other meson types is treated only approximately, i.e. in analogy to
the $\pi+\pi$ collisions by means of  mass-scaled cross sections.

We note, that another important source of the photons is the {\it bremsstrahlung in
meson+baryon collisions} (cf. Ref.~\cite{Linnyk:2013wma}). As we have
shown above, the SPA gives a  good approximation to the exact rates,
if we use the correct invariant energy in the hadronic sub-process
$s_2=s-q_0 \sqrt{s}$ and a realistic model for the differential
cross section of the subprocess, i.e. for the elastic scattering of
mesons on baryons. The cross sections for the meson+baryon elastic
scatterings (implemented within the PHSD transport approach) have
been previously adjusted to the data differentially in energy and
angular distribution.  Thus we evaluate the photon production in
the processes $m+B\to m+B+\gamma$ in the PHSD by using realistic
elastic scattering cross sections taken at the correct invariant
energy $\sqrt{s_2}$ in the scope of the improved SPA.

\subsection{The LPM effect }
\label{sect:LPM}

The radiation of photons by charged particles  is modi\-fied in the
medium compared to the vacuum. One of such medium effects is caused
by the absence of well-defined incoming and outgoing asymptotic
states due to the multiple scattering of particles in a strongly
interacting environment. If the subsequent scatterings occur within
the time necessary for photon radiation $\tau_\gamma \sim 1/q_0$,
then the amplitudes for the emission of photons before and after the
charged particle  scattering have to be summed coherently. The
effect of this destructive interference on the photon spectrum  by
electrons transversing a dense medium was first studied by Landau
and Pomeranchuk in Ref.~\cite{Landau:1953um} and Migdal in
Ref.~\cite{Migdal:1956tc}. Accordingly, the
Landau-Pomeranchuk-Migdal (LPM) effect modifies the spectrum of
photons produced in the medium in comparison to the incoherent sum
of emissions in quasi-free scatterings, leading especially to a
suppression of the low energy photons because the formation time of
the photon $\tau_\gamma$ is proportional to the inverse photon
energy $1/q_0$. In particular, the LPM effect regularizes the
$1/q_0$ divergence of the quasi-free bremsstrahlung spectra. The LPM
suppression and the induced thermal mass of the medium quanta (the
dielectric effect) together ensure that the photon spectrum is
finite in the limit $q_0\to 0$.

The importance of the LPM effect for the case of dilepton and photon
production from QCD systems was shown in
Refs.~\cite{Cleymans:1992kb,Cleymans:1992je,Knoll:1993ic}.
The magnitude of the LPM suppression is governed by the average time
between the collisions $\tau$, which in turn is given by the inverse
scattering length $a$ or by the inverse average spectral width of the
particles~$\Gamma$:
\be \tau=\frac{1}{a} \approx \frac{1}{\Gamma}. \ee
The LPM suppression is more pronounced in case of small $\tau$, i.e.
for high reaction rates. Thus we expect it to be important for the
emission of photons from the strongly-interacting quark-gluon plasma
(sQGP) as created in the early phase of the heavy-ion collision.
Indeed, it was shown in Refs.~\cite{BrCa11,Cassing:2013iz} in the
scope of the DQPM  that the average collision time of partons is as
short as $\tau\approx 2\!-3$~fm/c for temperatures in the range
$T\!=\!1\!\!-\!2 \ T_c$, where $T_c \approx 158$~MeV is the
deconfinement transition temperature. In comparison, the average
time between pion collisions in a thermalized pion gas at
temperatures $T<T_c$ is above 10~fm/c, see
Ref.~\cite{Cleymans:1992kb}.

%
%
\begin{figure}
\includegraphics[width=0.45\textwidth]{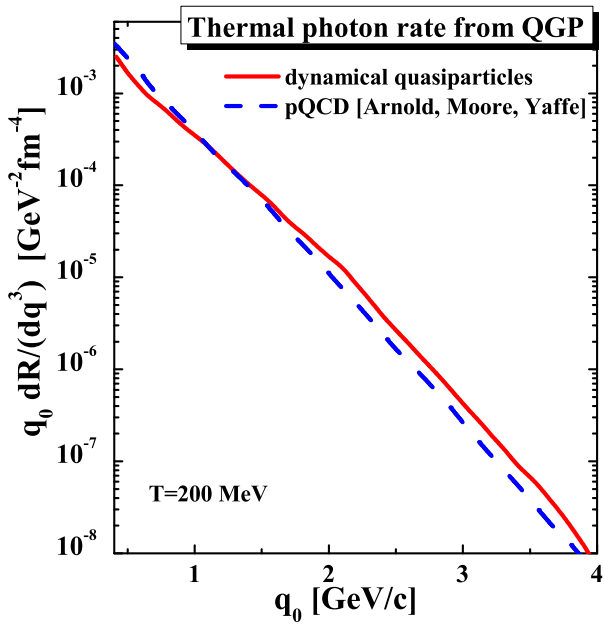}
\caption{(Color on-line) Invariant rate of photons produced from the
strongly-interacting quark-gluon plasma consisting of massive broad
quasi-particle quarks and gluons (red solid line). The leading-order
pQCD rate (blue dashed line) from Ref.~\protect{\cite{Arnold:2001ms}}
is shown for comparison.\label{amy} }
\end{figure}

The photon production in the sQGP proceeds by the interactions of
dressed quarks and gluons through quark annihilation and  gluon
Compton scattering processes $q+{\bar q}\rightarrow g+\gamma $,
$q({\bar q})+g \rightarrow q({\bar q})+\gamma$ as well as through
the quark bremsstrahlung and off-shell parton decays. In the
strongly interacting QGP (and in PHSD), the gluon and quark
propagators differ significantly from the non-interacting
propagators such that bare production amplitudes can no longer be
used~\cite{Cassing:2007nb,Cassing:2008nn}. The off-shell quarks and
gluons have finite masses and widths, which parametrize a resummed
interaction of the QGP constituents. The quark off-shellness leads
to higher twist corrections ($\sim m_q^2/s,m_q^2/t,m_q^2/u$) to
photon and dilepton production cross
sections~\cite{olena2010,Linnyk:2004mtLinnyk:2006mv}. These
corrections are small in hard hadron scattering at high
center-of-mass energy $\sqrt{s}>10$~GeV, but become substantial for
photon production in the sQGP, where the characteristic $\sqrt{s}$
of parton collisions is of the order of a GeV (cf. r.h.s. of  Fig.~\ref{dNdS}).
Using the cross sections for photon radiation by dressed quarks and
gluons in the processes $q \bar q \to g \gamma$ and $qg\to q \gamma$
as derived in Ref.~\cite{olena2010}, we have calculated the rates of
photons produced in a thermalized strongly interacting QGP.
Fig.~\ref{amy} presents the invariant rate of photons produced from
the strongly-interacting quark-gluon plasma   at the temperature
$T=200$~MeV (red solid line). The leading-order Log-resummed
perturbative QCD rate (blue solid line) from Arnold, Moore and Yaffe
(AMY rate) is taken from Ref.~\protect{\cite{Arnold:2001ms}} and
shown for comparison. One observes a qualitative agreement between
the results of both models although the degrees-of-freedom and their
couplings are different. We mention that photon rates recently
calculated at the NLO in  perturbative
QCD~\cite{Ghiglieri:2013gia,Vujanovic:2014xva,Ghiglieri:2014vua}
also approximately are in line with those presented in
Fig.~\ref{amy}.

Let us now quantify the magnitude of the LPM effect on the spectrum
of photons radiated from the QGP as calculated within the PHSD. The
coherent photon production rate - taking into account the LPM effect
- differs from the non-coherent cross section by a suppression
factor, which generally depends on the photon energy, temperature
and the interaction strength of the constituents. The coherent
photon emission rate was derived in Ref.~\cite{Cleymans:1992kb} for
an elastically interacting pion gas in the soft photon approximation
for the photon radiation amplitudes. The authors of
Ref.~~\cite{Cleymans:1992kb} used the same
method for the calculation of the photon emission over the whole
trajectory of the charged particle as was adopted in the original
work by Migdal in Ref.~\cite{Migdal:1956tc}. After averaging over
the times between collisions $\xi$, assuming an exponential
distribution
\be \frac{d W}{d \xi} = a e^{-\xi a}, \label{scattimes} \ee
the coherent photon emission rate was found to be
\be \frac{dR}{dq^3}=N\frac{2 \alpha_{_{EM}}}{(2 \pi)^2} \left< v^2
\frac{(1-\cos^2\Theta)}{a^2+q_0^2(1-v\cos \Theta)^2} \right> ,
\label{redl} \ee
where the brackets $<.>$ stand for an average over the velocities
after the scattering ($v$, $\cos \Theta$), while $N$ is the number
of scatterings and $\alpha_{EM}\approx 1/137$.  A realistic
parametrization of the data was used for the pion elastic scattering
cross section (cf. section~\ref{sect:brems}), but the scattering was
assumed isotropic. The incoherent rate is obtained from (\ref{redl})
in the limit $a=0$.

An analytical form of the coherence factor was obtained in
Ref.~\cite{Knoll:1993ic} in the model of hard scattering centers,
using a quantum mechanical approach to coherently sum the photon
amplitudes from all the scatterings. In the thermal medium the
spatial distribution of the scattering centers is assumed random.
Consequently, the function (\ref{scattimes}) naturally arises in
this model for the distribution of times between collisions by a
direct calculation of the two-particle correlation function. The
quenching factor in the dipole limit ($\vec q=0$) was found as
\be (G(q_0 \tau))^2 = \left( \frac{(q_0 \tau)^2}{1+(q_0 \tau)^2}
\right)^2. \label{lenk} \ee
Although formula (\ref{lenk}) was obtained in a simple model, it is
useful because it correctly captures the dependence of the LPM
suppression on the average strength of the interaction given solely
by the mean-free-time between collisions $\tau$ in the assumption of
isotropic collisions.

The perturbative interaction of quarks and gluons is dominated by
small scattering angles due to the massless particle exchange in the
$t$-channel diagrams. In this case the coherence factor for the
quark system in the limit of small scattering angles was obtained in
Ref.~\cite{Cleymans:1992je}. However, up to now the LPM effect in
case of a strongly interacting QGP with dressed broad
quasi-particles has not been evaluated.  The elastic scattering of
dressed quarks in the PHSD is not dominated by the $t\to0$ pole as
in the perturbative case since the gluon mass (of order 1 GeV) acts
as a regulator in the amplitude. Accordingly, the angular
distribution for quark-quark scattering is closer to an isotropic
distribution for low or moderate $\sqrt{s}$ in accordance with the
model assumptions of Ref.~\cite{Knoll:1993ic} such that the
expression (\ref{lenk}) should apply as an estimate of the LPM
suppression for the photon emission within the PHSD.

\begin{figure}
\includegraphics[width=0.45\textwidth]{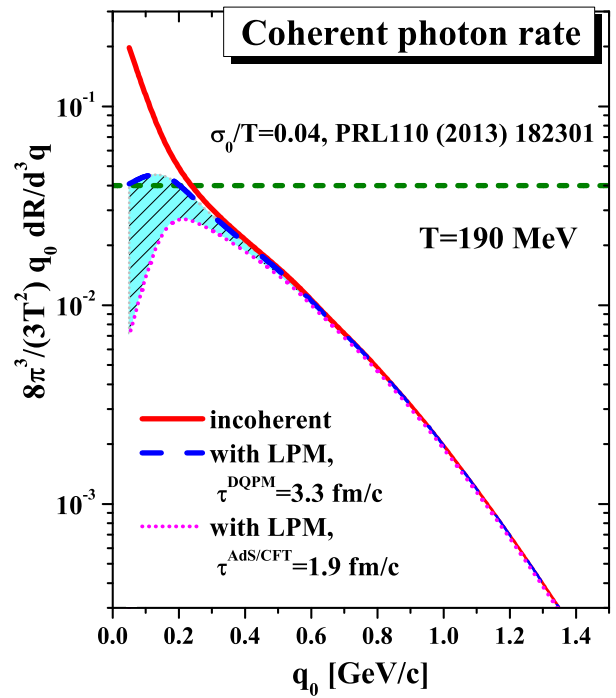}
\caption{(Color on-line) Incoherent invariant photon production rate
from the strongly-interacting quark-gluon plasma consisting of
massive broad quasi-particle quarks and gluons (red solid line)
scaled by $8\pi^3/(3 T^2)$ in order to match the electric
conductivity for $q_0\to0$ (cf. Eq. (\protect{\ref{ratecond}})). The
blue dashed line and the magenta dotted line show the coherent rates with
the two assumptions for the average time between the collisions
$\tau$ -- from the DQPM model and from the AdS/CFT correspondence.
\label{LPM} }
\end{figure}

In Fig.~\ref{LPM} we show the photon emission rate in a QGP at the
temperature $T$=190 MeV as calculated in the PHSD as an incoherent
sum of the photon emission in quark and gluon scatterings (red solid
line). The blue dashed line gives the same rate with the quenching
factor (\ref{lenk}) applied using $\tau(T)=1/\Gamma(T)\approx 3.3
$~fm/c from the DQPM (for $T$= 190 MeV).  We observe that the
suppression in comparison to the incoherent rate is visible only for
photon energies  $q_0<0.4$~GeV. For an estimate of the upper limit
on the LPM suppression we employ the relaxation time approximation
for the ratio of the shear viscosity over entropy density $\eta/s$
which gives $\eta/s \approx $0.14 at $T$=190 MeV in
DQPM~\cite{Ozvenchuk:2012fn,Ozvenchuk:2012kh}. The lowest bound as
conjectured within the AdS/CFT correspondence is $\eta/s=1/(4 \pi)
\approx 0.08$. In the relaxation time approximation this corresponds
to a lower value of $\tau \approx 1.9$~fm/c. The coherent photon
rate in this case is given by the magenta dotted line and shows a
peak in the photon rate for $q_0 \approx$ 0.2 GeV.

In order to further clarify the strength of the LPM suppression of
the photon emission in the sQGP, we use the knowledge of the
electric conductivity $\sigma_0(T)$ of the sQGP from the
DQPM~\cite{Cassing:2013iz} which is roughly in line with more recent
results from lattice QCD. For details of the latter analysis on the
electric and magnetic response of the sQGP we refer the reader to
Ref. \cite{Steinert:2013fza}. We recall that the photon emission
rate from a thermal medium is controlled by
$\sigma_0$~\cite{Yin:2013kya} via the relation,
\be \frac{\sigma_0}{T}=\frac{8 \pi^3}{3 T^2} \lim _ {q_0\to 0}
\left( q_0 \frac{d R}{ d^3 q} \right), \label{ratecond} \ee
where $T$ is the temperature of the system, $q_0$ is the photon
energy and $\vec q$ is the photon momentum.
%
%
Using the number for $\sigma_0/T$  from the PHSD at the temperature
of $T=190$~MeV from Ref.~\cite{Cassing:2013iz}, we obtain a limiting
value for the scaled photon emission rate of $0.04$ for $q_0\to0$
according to formula (\ref{ratecond}) (green short dashed line in
Fig.~\ref{LPM}). The blue dashed line in Fig.~\ref{LPM} -- the
estimate of the rate based on  formula (\ref{lenk}) and the DQPM
average spectral width of the quarks/antiquarks -- indeed approaches
the limiting value of 0.04 as given by the kinetic calculations of
the electric conductivity in
Refs.~\cite{Cassing:2013iz,Steinert:2013fza}.

Taking into account some uncertainty in the determination of $\tau$
and the expression (\ref{lenk}), we conclude from Fig.~\ref{LPM} and
analogous calculations at different temperatures that the LPM effect
influences the photon production from the QGP at photon energies
below $q_0 \approx 0.4$~GeV, but is negligible for higher photon
energies. We note in passing that the suppression of the photon
spectrum in the hadronic phase is much smaller due to the lower
interaction rate, i.e. longer interaction time $\tau$.

\subsection{Additional meson-nucleon processes}
\label{sect:22}

In the present work we incorporate into the PHSD approach
additionally the $2\to2$ processes  $V + N \rightarrow N + \gamma$,
where $V$ stands for a vector meson while $N$ denotes a proton or
neutron. These processes are the baryonic counterparts to the
mesonic $2\to2$ reactions $\pi+\rho/\pi\to\gamma+\pi/\rho$. We here
consider the interaction of nucleons with the mesons $V=\rho$,
$\phi$, $\omega$, taking into account the various possible charge
combinations, e.g. $\rho^0+p\to \gamma+p$, $\rho^-+p\to\gamma+n$,
$\rho^+ + n\to \gamma +p$, etc. Additionally, we take into account
the photon production in the decay of the $\Delta$-resonance,
$\Delta\to N+\gamma$.

In order to evaluate the probabilities for photon production in the
collisions of vector mesons with  nucleons, we use the inverse
processes $\gamma+N\to\rho+N$, $\gamma+N\to\phi+N$,
$\gamma+N\to\omega+N$ (controlled by data) and employ detailed
balance to obtain the differential cross sections for the processes
$\rho+N\to\gamma+N$, $\phi+N\to\gamma+N$, $\omega+N\to\gamma+N$.

We recall that the detailed balance formula  reads:
\bea \sigma(NV\to \gamma N) & = & \frac{g_\gamma}{g_V} \frac{
p^{*2}_{N\gamma}}{p^{*2}_{NV}} \  \sigma(\gamma N \to NV),
 \eea
where $g_\gamma=2$ and $g_V=3$ are the spin degeneracy factors of
the photon and the vector meson $V$, $p^*_{N\gamma}$ is the
center-of-mass momentum in the $N+\gamma$ system, and $p^*_{NV}$ is
the center-of-mass momentum in the $N+V$ system.

The cross sections for the exclusive photo-production of $\rho$,
$\phi$ and $\omega$ vector mesons on the nucleon have been measured
by the Aachen-Berlin-Bonn-Hamburg-Heidelberg-Munich (ABBHHM)
Collaboration and published in Ref.~\cite{ABBHHM:1968aa}. In the
same work also parametrizations for these cross section have been
given that are based on the vector-meson-dominance model with a
non-relativistic Breit-Wigner (BW) spectral function for the
$\rho$-meson. Later, the fits have been updated in Ref.
\cite{Effenberger:1999ay} using  relativistic BW spectral functions
for  $\rho$, $\omega$ and $\phi$ mesons.

The total cross sections -- fitted in Ref.~\cite{Effenberger:1999ay}
to the data from Ref.~\cite{ABBHHM:1968aa} -- are given by
\be \sigma(\gamma N\to VN) =\frac{1}{p^*_{N\gamma} s} \int d\mu
|M_V|^2 p^*_{NV} A_V(\mu), \label{cros} \ee
where the mass of the vector meson is distributed by the spectral
function $A_V(\mu)$:
\be A_V (\mu) = \frac{2}{\pi} \frac{\mu ^2 \Gamma (\mu)}{
(\mu^2-M_i^2)^2 + \mu^2 \Gamma^2(\mu) }, \ee
with $M_i$ denoting the pole mass of the meson. The matrix elements
for $\gamma + N \to V + N$  are parametrized as
 \bea
|M_\rho|^2&\!=\!&0.16 \mbox{ mb GeV}^2, \nn
|M_\omega|^2&\!=\!&\frac{0.08 \ p_{VN}^{*2}}{2(\sqrt{s}-1.73 \mbox{\,
GeV})^2+p_{VN}^{*2}} \mbox{ mb GeV}^2, \nn
 |M_\phi|^2&\!=\!&0.004 \mbox{ mb
GeV}^2.
 \label{fits} \eea
The cross sections (\ref{cros}) with the parameters (\ref{fits}) are
consistent with the dynamics of vector mesons in the PHSD, where
also relativistic BW spectral functions for vector mesons are used
and propagated off-shell.

For the angular distribution of the $\rho$-meson production in the
process $\gamma+N\to N+\rho$, we follow  the suggestion of
Ref.~\cite{Effenberger:1999ay},
\be \frac{d \sigma}{dt} \sim \exp (B t), \label{param} \ee
with the photon-energy dependent parameter $B$ (fitted to the data):
$B=5.7$ for $q_0 \le 1.8$~GeV,
$B=5.43$ for $1.8 < q_0 \le 2.5$~GeV,
$B=6.92$ for $2.5 < q_0 \le 3.5$~GeV,
$B=8.1$ for $3.5 < q_0 \le 4.5$~GeV,
$B=7.9$ for $q_0 > 4.5$~GeV.

The data in Ref.~\cite{Effenberger:1999ay} have shown that the cross
section is dominated by the $t\approx 0$ region in line with the
physics assumptions of the vector dominance model (VDM) where the
process $\gamma + N \to V+N$ is described by the incident photon
coupling to the vector meson of helicity $\pm1$, which consequently
is scattered elastically by the nucleon (cf.
Refs.~\cite{Klopot:2013laa,Sakurai:1960ju,Chew:1957tf}).

Let us now briefly describe the modeling of the photon production in
the decays of the $\Delta$-resonance. The $\Delta \to N\gamma$ width
depends on the resonance mass $M_\Delta$, which is distributed
according to the $\Delta$ spectral function. Starting from the
pioneering work of Jones and Scadron \cite{Jones73}, a series of
models \cite{Wolf90,Krivor02,ZatWolf03} provided the mass-dependent
electromagnetic decay width of the $\Delta$ in relation to the total
width of the baryon. We employ the model of Ref.~\cite{Krivor02} in
the present calculations where the spectral function of the $\Delta$
resonance is assumed to be of relativistic Breit-Wigner form. We
adopt the "Moniz" parametrization \cite{Moniz} for the shape of the
$\Delta$ spectral function, i.e. the dependence of the width on the
mass $\Gamma^{tot}(M_\Delta)$.

\section{Comparison to data from heavy-ion collisions}
\label{sect:results}

Before presenting the results on photon spectra from heavy-ion
collisions we stress again that the PHSD approach so far has
provided a consistent description of the bulk properties of
heavy-ion collisions -- rapidity spectra, transverse mass
distributions, azimuthal asymmetries of various particle species --
from low SPS up to  LHC
energies~\cite{CasBrat,BrCa11,Konchakovski:2014fya}. Furthermore,
dilepton production from hadronic and partonic sources has been
calculated at SPS, RHIC and LHC
energies~\cite{Linnyk:2011hz,Linnyk:2011vx,Linnyk:2012pu} and
successfully compared to the available dilepton data
\cite{Bratkovskaya:2014mva}. Accordingly, the global dynamics of the
bulk matter, its collective flow as well as the electromagnetic
emissivity in heavy-ion collisions appear to be well under control.
Here we extend our previous photon studies in
Refs.~\cite{Linnyk:2013hta,Linnyk:2013wma} at the top RHIC energy
also to the top SPS and LHC energy incorporating the improved cross
sections as described above.

The inclusive photon yield as produced in $p+p$ and $A+A$ collisions
is divided into ``decay photons" and ``direct photons". {\it Decay
photons } -- which constitute the major part of the inclusive
photons -- stem from the photonic decays of hadrons (mesons and
baryons). These decays occur mainly at later times and outside of
the active reaction region and therefore carry limited information
on the initial high-energy state. Consequently, it is attempted to
separate the decay photons from the inclusive yield (preferably by
experimental methods) and to study the remaining, {\em direct
photons}. One usually uses the ``cocktail" method to estimate the
contribution of the photon decays to the spectra and to the elliptic
flow $v_2$, which relies among others on the $m_T$-scaling
assumption and on the photon emission only by the finally produced
hadrons with the (momenta) distributions as in the final state.
Depending on the particular experimental set-up, different
definitions of the decay photons are applied: whether only the
decays of $\pi^0$- and $\eta$-mesons are attributed to the decays
photons or also the decays of the less abundant and shorter-living
particles $\eta^\prime$, $\omega$, $\phi$, $a_1$ and the
$\Delta$-resonance. Indeed, the determination of the latter
contributions (in particular, $a_1$, $\Delta$) by experimental
methods is questionable, because of the emission during the
absorption and regeneration in the initial interacting phase.
Therefore, the theoretical understanding of the decay photon
contributions to the inclusive spectrum is important. Especially
when analyzing simultaneously various measurements at different
energies and within different experimental settings. One can
consider theoretically {\em decays} photons from the following
processes:
\bea & \pi^0 \to \gamma+ \gamma, \mbox{\hspace{0.4cm}} \eta \to
\gamma + \gamma, \mbox{\hspace{0.4cm}} \eta^\prime \to \rho +
\gamma, &
\nonumber \\
& \omega  \to  \pi^0 + \gamma, \mbox{\hspace{0.4cm}} \phi \to \eta +
\gamma, \mbox{\hspace{0.4cm}}
 a_1  \to  \pi + \gamma, \mbox{\hspace{0.4cm}} \Delta  \to
\gamma+N, \nonumber & \eea
where the parent hadrons are produced in baryon-baryon ($BB$),
meson-baryon ($mB$) or meson-meson ($mm$) collisions in the course
of the heavy-ion collision.

The following contributions to the {\em direct} photons have been
identified so far:
\begin{itemize}
\item The photons at large transverse
momentum $p_T$, so called ``prompt" or ``pQCD" photons, are produced
in the initial hard $N+N$ collisions and the from the jet
fragmentation, which are well described the perturbative QCD (pQCD)
calculations. The latter, however, might be modified in $A+A$
contrary to $p+p$ due to the modification of the parton
distributions (initial state effect) the parton energy loss in the
medium (final state effect). In the $A+A$ collisions at large $p_T$,
there may also be contributions from the induced
jet-$\gamma$-conversion in the QGP and the jet-medium photons from
the scattering of hard partons with thermalized partons
$q_{hard}+q(g)_{QGP} \to \gamma + q(g)$, but these are relatively
small. The prompt photons are well modeled by using the perturbative
calculations and adjusting them to the high-$p_T$ region of the
observed direct photon yield.
\item After the deduction of the prompt photons from the direct
photon spectra, there is a significant remaining photon radiation at
$p_T<3$~GeV observed, which is dubbed as {\it ``thermal" photons}.
The low-$p_T$ photons are emitted by the various partonic and
hadronic sources as listed below:
\begin{enumerate}
\item
Photons are radiated by  quarks in the interaction with other quarks
and gluons, such as in the reactions $$q+\bar{q}\rightarrow
g+\gamma, \hspace{2cm} q/\bar{q}+g\rightarrow q/\bar{q}+\gamma.$$ In
addition, photon production in the bremsstrahlung reactions
$q+q/g\to q+q/g+\gamma$ are possible.
\item
All colliding hadronic charges (meson, baryons) can also radiate
photons by the bremsstrahlung processes:
$$
\hspace{1.6cm} m+m\to m+m+\gamma \mbox{\hspace{0.7cm} }  m+B\to
m+B+\gamma .
$$
\item Additionally, the photons can be produced in binary
meson+meson and meson+baryon collisions. We consider within the PHSD
the direct photon production in the following $2\to2$ scattering
processes:
\bea & \pi + \pi \rightarrow \rho + \gamma , \mbox{\hspace{0.9cm}}
\pi + \rho \rightarrow \pi + \gamma , & \nonumber
\\ &
 V+N \to \gamma+N, &  \nonumber
\\  & \mbox{where } V=\rho, \
\phi, \  \omega, \mbox{ and }N=n,p, & \nonumber \eea
accounting for all possible charge combinations.
\end{enumerate}
\end{itemize}

\subsection{Direct photon spectra from SPS to LHC}

\begin{figure}
\includegraphics[width=0.45\textwidth]{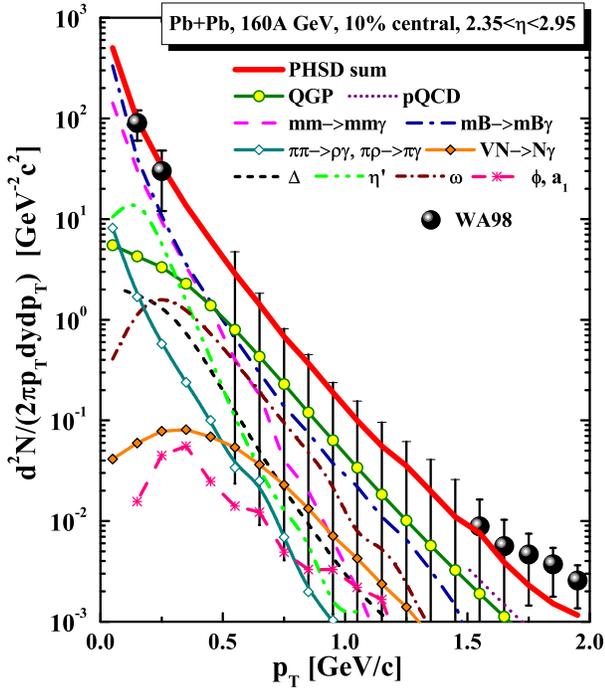}
\caption{(Color on-line) Comparison of the PHSD calculations to the
data of the WA98 Collaboration
from~\cite{Aggarwal:2000thAggarwal:2000ps}.  In comparison to the
original HSD study~\cite{Bratkovskaya:2008iq}: (i) the meson+baryon
bremsstrahlung (blue dash-dotted line), $\Delta$ decays (black
short-dashed line) and the photons from QGP (green line with round
symbols) are added (ii) and the meson+meson bremsstrahlung is now
calculated beyond the SPA (magenta dashed line). The black line with
diamond symbols labeled as ``other" includes: $\omega$, $\eta'$,
$\phi$ an d $a_1$-meson decays, binary channels
$\pi+\rho/\pi\to\pi/\rho+\gamma$ and $N+V\to N+\gamma$. \label{WA98}
}
\end{figure}

We start with the system Pb+Pb at $\sqrt{s_{NN}}$ = 17.3 GeV, i.e.
at the top SPS energy. Fig.~\ref{WA98} shows the comparison of the
PHSD calculations to the data of the WA98 Collaboration
from~\cite{Aggarwal:2000thAggarwal:2000ps} for 10\% centrality in
the pseudorapidity interval $2.35 < \eta < 2.95$. In addition to the
sources, which have been incorporated in the original HSD
study~\cite{Bratkovskaya:2008iq}, the meson+baryon bremsstrahlung,
$VN\to N \gamma$, $\Delta\to N\gamma$ decay and the QGP channels are
added. Compared to the earlier results of
Ref.~\cite{Bratkovskaya:2008iq}, the description of the data is
further improved and the conclusions remain unchanged: the
bremsstrahlung contributions are essential for describing the data
at low $p_T$.  This interpretation is shared by the authors
of Refs.~\cite{Dusling:2009ej,Haglin:2003sh,Liu:2007zzw}, who also
stressed the importance of the meson+meson bremsstrahlung in view of
the WA98 data using  hydrodynamical or fireball models. Note that
the photon contribution from the QGP is practically negligible for
low $p_T$ and reaches at most 25\% at $p_T>0.5$~GeV.

\begin{figure}
\includegraphics[width=0.45\textwidth]{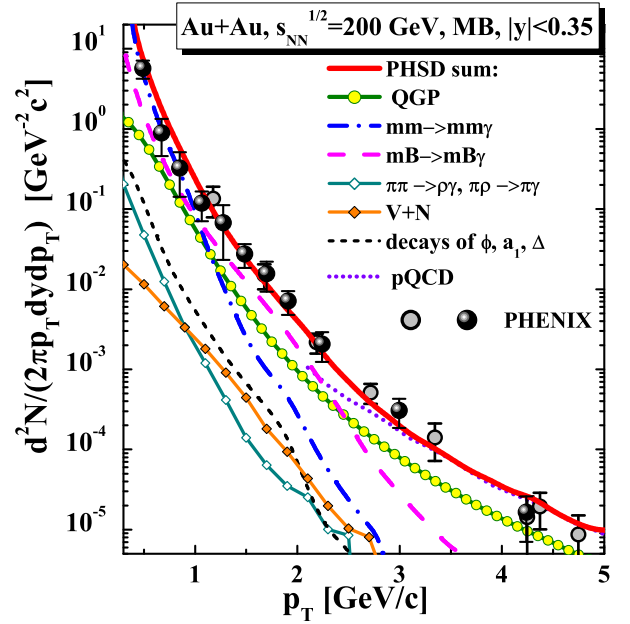}
\caption{(Color on-line) PHSD results for the spectrum of direct
photons produced in minimum bias Au+Au collisions at
$\sqrt{s_{NN}}=200$~GeV as a function of the transverse momentum
$p_T$ at mid-rapidity $|y|< 0.5$. The data of the PHENIX
Collaboration are taken from Refs.
\protect{\cite{Adare:2008ab,Adare:2014fwh}}. For the individual
lines see the legend in the figure. \label{spectrarhic} }
\end{figure}

\begin{figure*}
\includegraphics[width=0.9\textwidth]{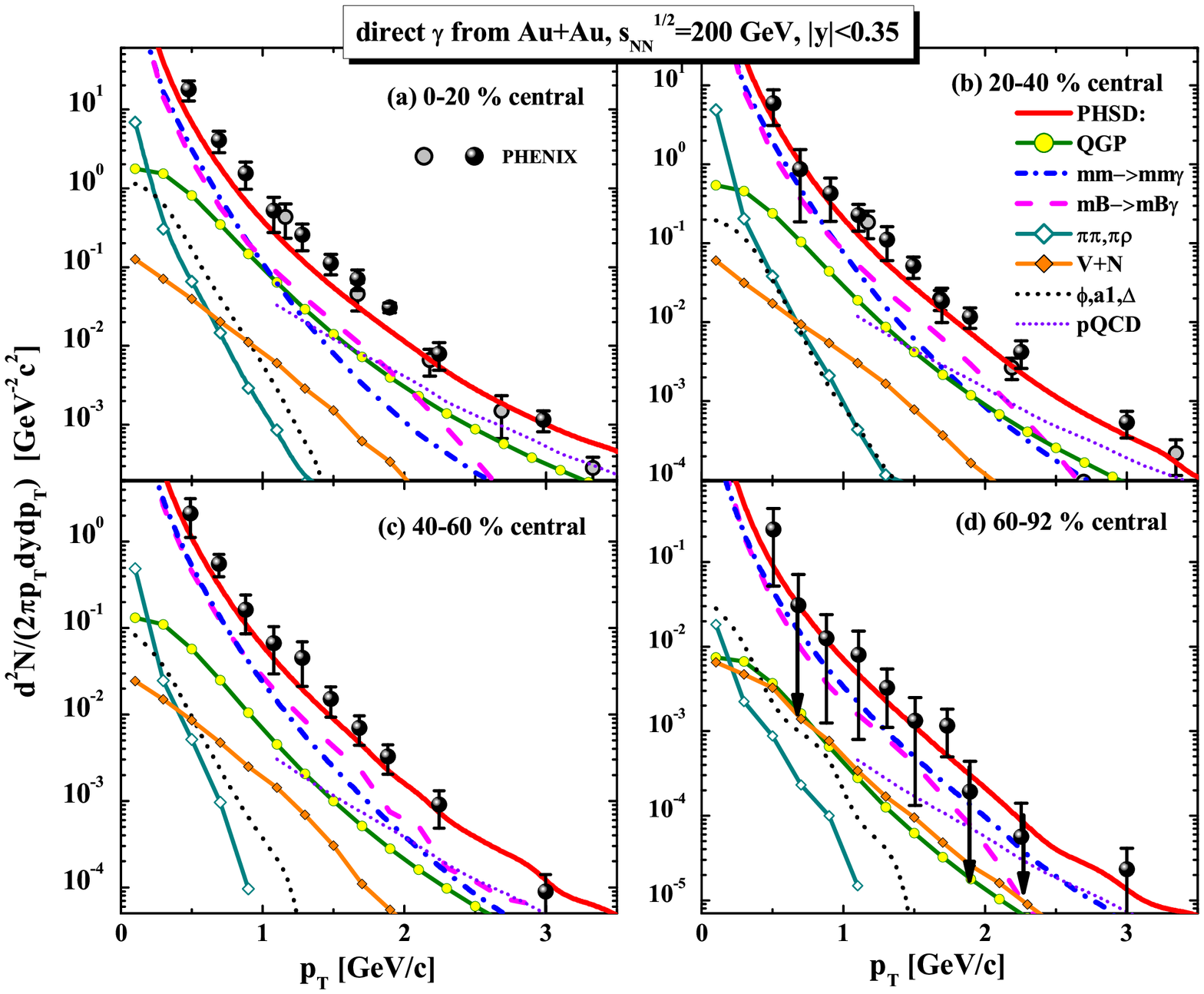}
\caption{(Color on-line) Contribution of the photon production in
the two-to-two $\rho$+nucleon interaction (orange dashed lines) to
the total direct photon spectra (red lines) at the top RHIC energy
for different centralities. The dominant sources are the photons
from the QGP and from the hadronic two-to-three bremsstrahlung
processes. The PHENIX data are from Refs.
\protect{\cite{Adare:2008ab,Adare:2014fwh}}. \label{222} }
\end{figure*}

We now step on to the top RHIC energy of $\sqrt{s_{NN}}$ = 200 GeV
and update our results on the differential photon spectra for the
system Au+Au.  The results for the direct photon spectrum as a sum
of partonic as well as hadronic sources for the photons produced in
minimum bias Au+Au collisions  is presented in
Fig.~\ref{spectrarhic} as a function of the transverse momentum
$p_T$ at mid-rapidity $|y|< 0.5$.  This observable has been
calculated within the PHSD
earlier~\cite{Linnyk:2013hta,Linnyk:2013wma}, but has to be
revisited in the present work since we have incorporated additional
photon production channels such as the binary baryon+meson
collisions $V+N\to N+\gamma$, $\Delta \to N \gamma$ and improved the
calculation of the $m+m\to m+m + \gamma$ bremsstrahlung channel
based on the OBE model results which go beyond the soft-photon
approximation. Indeed, a direct comparison of Fig.~\ref{spectrarhic}
to our previous results for the direct photon spectra at RHIC (cf.
Fig.~1 in Ref.~\cite{Linnyk:2013wma} and Fig.~3 of
Ref.~\cite{Linnyk:2013hta}) shows that the data seem to favor the
OBE model employed here in comparison to the SPA with the simple
constant isotropic cross section employed earlier. In particular the
agreement with the data is improved in the high-$p_T$ range.

We recall that
in our calculations the direct photon spectrum has the following
contributions: photon bremsstrahlung in meson+meson $m+m\to
m+m+\gamma$ (blue dash-dotted line) and meson+baryon collisions
$m+B\to m+B+\gamma$ (magenta dashed line); photon production in the
QGP in the processes $q+{\bar q} \to g+\gamma$, and $q({\bar q})+g
\to q({\bar q})+\gamma$ (green line with yellow circles); the
reactions $\pi+\rho\to\pi+\gamma$ and $\pi+\pi\to \rho+\gamma$,
(cyan line with open symbols); the meson+baryon reactions $V+N\to
N+\gamma$ (orange line with filled symbols); decays of $\phi$ and
$a_1$ mesons and of $\Delta$-baryons (black short-dashed line);
the photon production in the initial hard collisions ("pQCD")
is given by the hard photon yield in p+p collisions scaled with the
number of collisions $N_{coll}$ (violet dotted line). The measured
transverse momentum spectrum $dN/dp_T$ (given by the filled circles)
is reproduced well by the sum of all partonic and hadronic sources
 (red solid line).

\begin{figure}
\includegraphics[width=0.45\textwidth]{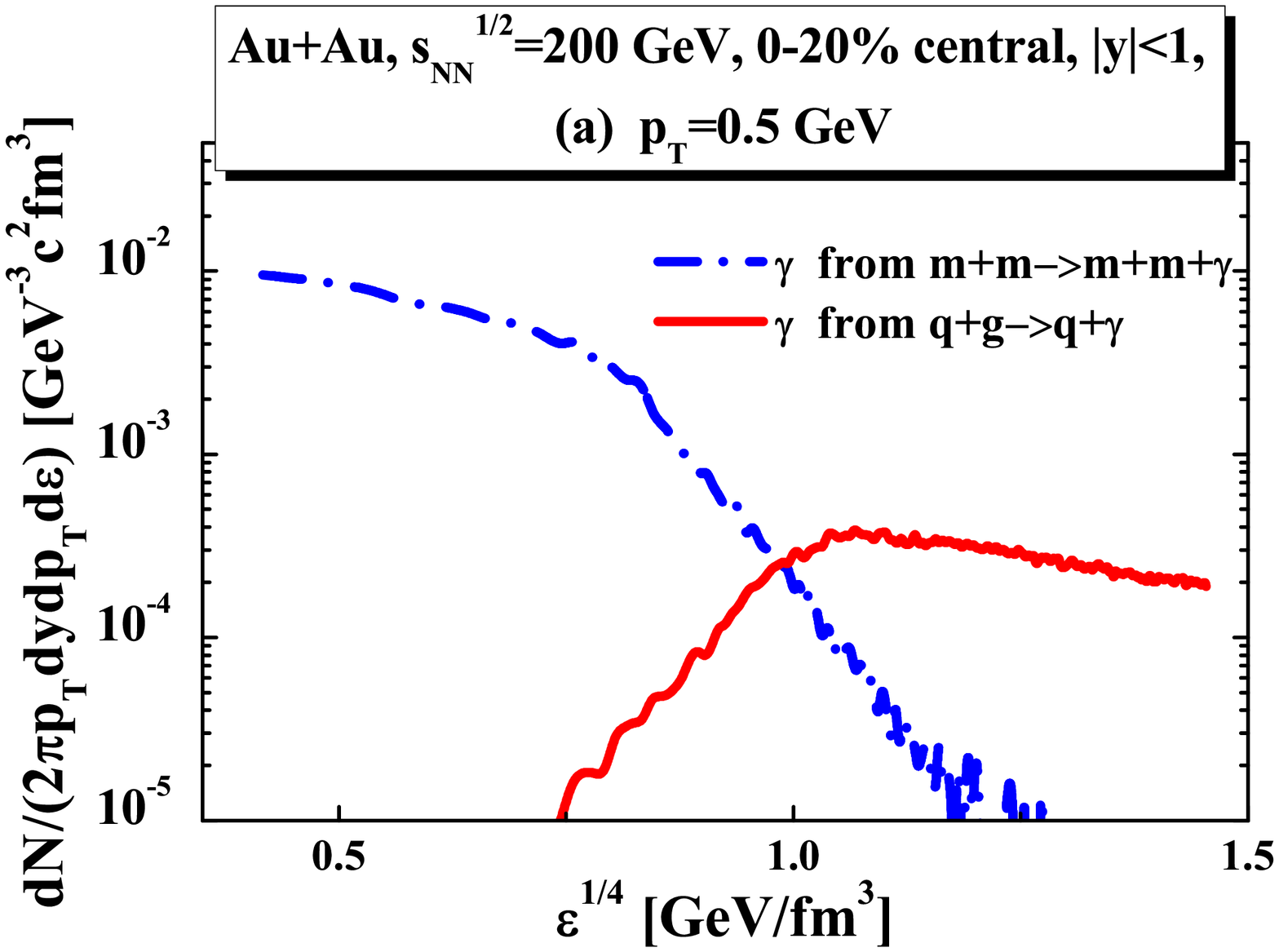}
\includegraphics[width=0.45\textwidth]{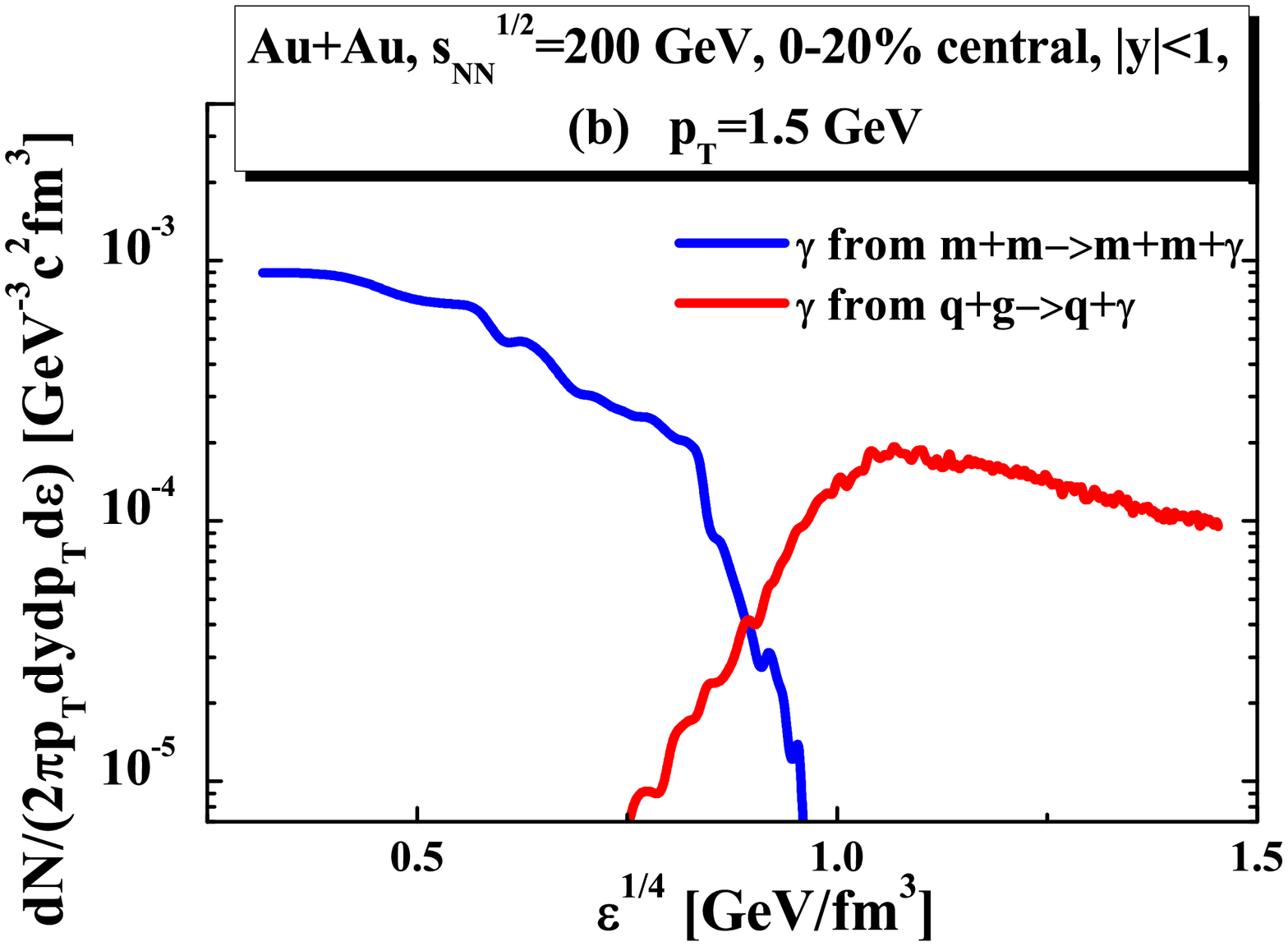}
\caption{(Color on-line){ Top:} Yield of photons with the transverse
momentum  $p_T=0.5$~GeV at mid-rapidity produced in 0-20 \% most
central $Au+Au$ collisions as functions of the approximate local
``temperature" (the fourth-root of the energy density) from the PHSD
from meson-meson bremsstrahlung (dash-dotted lines) and gluon
Compton scattering (solid lines). \label{Evol05} { Bottom: } Same as
in the top panel for photons with the transverse momentum
$p_T=1.5$~GeV.
 \label{Evol15}  }
\end{figure}

As we have previously stressed in Ref.~\cite{Linnyk:2013wma}, the
centrality dependence of the photon spectra carries additional
information, which can be used to disentangle the hadronic and
partonic sources of the photon production. The recent measurements
by the PHENIX Collaboration~\cite{Adare:2014fwh} confirm the
predictions within the PHSD from Ref.~\cite{Linnyk:2013wma}.  The
centrality dependence of the integrated thermal photon yield in PHSD
was found to scale as $N_{part}^\alpha$ with the exponent
$\alpha \approx 1.5$, which is in a good agreement with the most recent
measurement of $\alpha=1.48\pm0.08\pm0.04$ by the PHENIX
Collaboration~\cite{Adare:2014fwh}. We compare our calculations and
the data in Fig.~\ref{222}. In the present work we have added the
cross sections of the processes $V+N\to \gamma +N$ by exploiting
detailed balance from the cross sections (\ref{cros}) and the photons
from  the
$\Delta$-Dalitz decays. The contributions of these channels to the direct
photon yield in the $Au+Au$ collisions at $\sqrt{s_{NN}}=200$~GeV at
four different centralities are explicitly shown in Fig.~\ref{222}, where
the recent measurements by the PHENIX
Collaboration~\cite{Adare:2014fwh} are compared to the PHSD spectra, too.
It can be seen in Fig.~\ref{222} that the inclusion of the photon
production from the
$V+N$ reactions does not enhance the total direct photon yield much
at all four centralities. We find that partonic channels and
bremsstrahlung photon production ``overshine" the $2\to2$ reactions,
both meson+meson and (vector) meson+baryon channels.

The modification of vector meson spectral functions in the nuclear medium is an
interesting phenomenon, which has been found essential for the
understanding of dilepton production at  SPS and RHIC
energies~\cite{Linnyk:2011hz,Linnyk:2011vx}. Accordingly, there
should be also an effect of the in-medium modification of vector-meson
spectral functions on the  photon production in the $V+N$ reactions.
However, since the $V+N$ channel does not dominate the spectrum at
any photon energy, we conclude that this effect cannot be resolved
in the total direct photon spectra at the present level of the data
accuracy at RHIC; this has been explicitly shown earlier in
Ref.~\cite{Bratkovskaya:2008iq} for the top SPS energy.

Next, let us investigate  the photon production across the phase
transition in the heavy-ion collision to check whether the observed
yield of direct photons is produced dominantly in some particular
region of the energy-density or in some particular phase of matter.
Fig.~\ref{Evol05} shows the yield of photons produced at
mid-rapidity in 0-20 \% most central $Au+Au$ collisions at
$\sqrt{s_{NN}}$ = 200 GeV  as functions of the approximate local
``temperature" (i.e. the fourth-root of the energy density) from the
PHSD. The top panel of Fig.~\ref{Evol05} presents the calculations
for  photons with the transverse momentum $p_T=0.5$~GeV, while the
bottom panel corresponds to  photons with a transverse momentum
$p_T=1.5$~GeV. We observe that the early, hot state does not
dominate the photon production in the QGP contrary to expectations
of the static thermal fireball model, where photon production is
roughly proportional to a power of the temperature ($\sim T^4$). The
integration over the dynamical evolution of the heavy-ion collision
leads to roughly the same contribution of the different energy
density regions since the rate decreases but the space-time volume
increases. The photon production in the hadronic phase is dominated
by the lower energies/temperatures because of the very long times
over which the produced hadrons continue to interact elastically,
which is accompanied by the photon bremsstrahlung in case of charged
hadrons.


\begin{figure}
\includegraphics[width=0.45\textwidth]{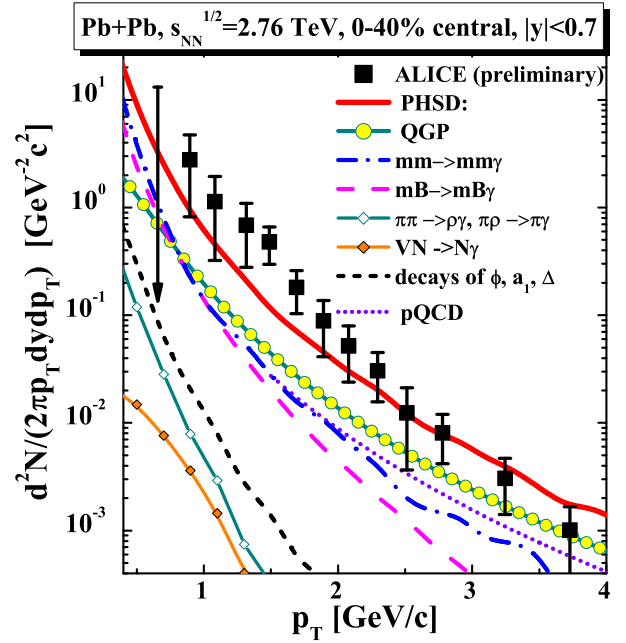}
\caption{(Color on-line) Yield of direct photons at mid-rapidity in
Pb+Pb collisions at the invariant energy $\sqrt{s_{NN}}=2.76$~TeV
for 0-40\% centrality within the PHSD in comparison to the
preliminary data from the ALICE
Collaboration~\protect\cite{Wilde:2012wc}. \label{spectralhc} }
\end{figure}

\begin{figure*}
\includegraphics[width=\textwidth]{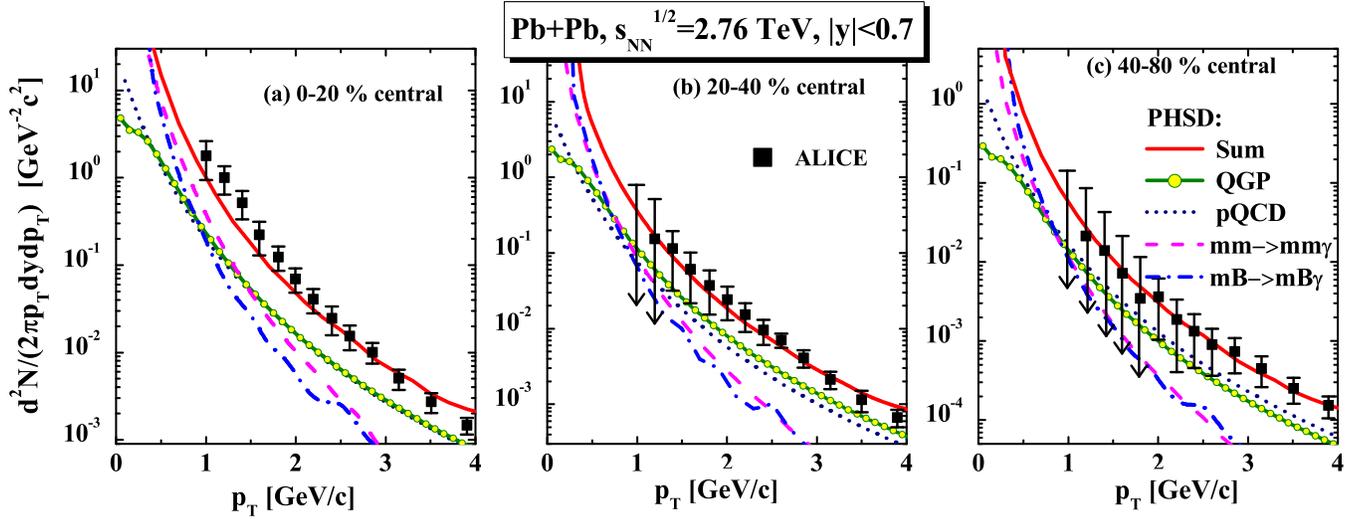}
\caption{(Color on-line) Transverse momentum spectra of direct
photons at mid-rapidity in Pb+Pb collisions at the invariant energy
$\sqrt{s_{NN}}=2.76$~TeV for the three centrality bins -- 0-20\%,
20-40\% and 40-80\% -- as predicted within the PHSD. The black
symbols show the very recently available data from the ALICE
Collaboration~\protect\cite{Adam:2015lda}. See the legend for the
individual contributions. \label{lhc_centr} }
\end{figure*}

We now increase the invariant energy $\sqrt{s_{NN}}$ by roughly
another factor of 14.  In Fig.~\ref{spectralhc} we show the direct
photon yield from PHSD in Pb+Pb collisions at the invariant energy
$\sqrt{s_{NN}}=2.76$~TeV for 0-40\% centrality. When comparing the
theoretical calculations to the preliminary data of the ALICE
Collaboration from Ref.~\cite{Wilde:2012wc}, we find a rather good
overall agreement with the data within about a factor of 2 in the
range of transverse momenta $p_T$ from 1 to 4 GeV. On the other
hand, the calculations tend to underestimate the preliminary data in
the low-$p_T$ region~\cite{Linnyk:2015nea}.

Furthermore, we provide predictions for the centrality-dependence of
the direct photon transverse momentum distributions at the LHC
energy in Fig.~\ref{lhc_centr}, since the differential investigation
in centralities and energies of the heavy-ion collisions will
provide crucial information for a clarification of the relative
importance of the contributing photon sources. Very recently, the
data of the ALICE Collaboration for the photon yield at these three
centralities has become available~\cite{Adam:2015lda}, we show them
as the black symbols in Fig.~\ref{lhc_centr}.

In conclusion, we have found that from SPS to LHC energies the
radiation from the sQGP constitutes less than half of the observed
number of direct photons for central reactions. The radiation from
hadrons and their interaction -- which are not measured separately
so far -- give a considerable contribution especially at low
transverse momentum. The dominant hadronic sources are the meson
decays, the meson-meson bremsstrahlung and the meson-baryon
bremsstrahlung. While the former (e.g. the decays of $\omega$,
$\eta$', $\phi$ and $a_1$ mesons) can be subtracted from the spectra
once the mesonic yields are determined independently by experiment,
the reactions $\pi+\rho\to\pi+\gamma$, $\pi+\pi\to \rho+\gamma$,
$V+N\to N+\gamma$, $\Delta \to N+\gamma$ as well as the meson-meson
and meson-baryon bremsstrahlung can be separated from the partonic
sources only with the assistance of theoretical models.

\begin{figure}
\includegraphics[width=0.45\textwidth]{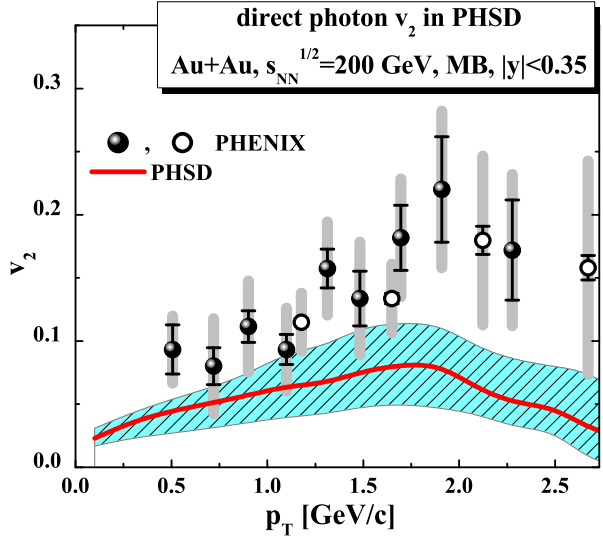}
\caption{(Color on-line) Elliptic flow $v_2$ versus transverse
momentum $p_T$ for the direct photons produced in minimum bias Au+Au
collisions at $\sqrt{s_{NN}}=200$~GeV calculated within the PHSD
(solid red line); the blue band reflects the uncertainty in the
modeling of the cross sections for the individual channels. The
 data of the PHENIX Collaboration are from
Ref.~\protect{\cite{PHENIX1}}. \label{rhicV2} }
\end{figure}

\subsection{Elliptic flow of direct photons}
\label{sect:v2}

The azimuthal momentum distribution of the emitted particles is
commonly expressed in the form of a Fourier series as
\be E\frac{d^3N}{d^3p}= \nonumber \frac{d^2N}{2\pi
p_Tdp_Tdy}\left(\! 1\!+\! \sum^\infty_{n=1} 2v_n(p_T) \cos
[n(\psi-\Psi_n)]\! \right)\!, \ee
where $v_n$ is the magnitude of the $n'$th order harmonic term
relative to the angle of the initial-state spatial plane of symmetry
$\Psi_n$ and $p=(E,\vec{p})$ is the four-momentum of the particle
under consideration. We here focus on the coefficients $v_2$ and
$v_3$ which implies that we have to perform event by event
calculations in order to catch the initial fluctuations in the shape
of the interaction zone and the event plane $\Psi_{EP}$. We
calculate the triangular flow $v_3$ with respect to $\Psi_3$ as
$v_3\{\Psi_3\} = \langle
\cos(3[\psi-\Psi_3])\rangle/\rm{Res}(\Psi_3)$. The event plane angle
$\Psi_3$ and its resolution $\rm{Res}(\Psi_3)$ are evaluated using
the hadron-hadron correlations at larger rapidities as described in
Ref.~\cite{{Adare:2011tg}} via the two-sub-events
method~\cite{Poskanzer:1998yz,Bilandzic:2010jr}.

We recall that the second flow coefficient $v_2$ carries information
on the interaction strength in the system -- and thus on the state
of matter and its properties -- at the space-time point, from which
the measured particles are emitted. The elliptic flow $v_2$ reflects
the azimuthal asymmetry in the momentum distribution of the produced
particles ($p_x$ vs $p_y$), which is a consequence of the
geometrical azimuthal asymmetry of the initial reaction region. If
the produced system is a weakly-interacting gas, then the initial
spatial asymmetry is not effectively transferred into the final
distribution of the momenta. On the contrary, if the produced matter has
the properties of a liquid, then the initial geometrical
configuration is reflected in the final particle momentum distribution.

\begin{figure}
\includegraphics[width=0.5\textwidth]{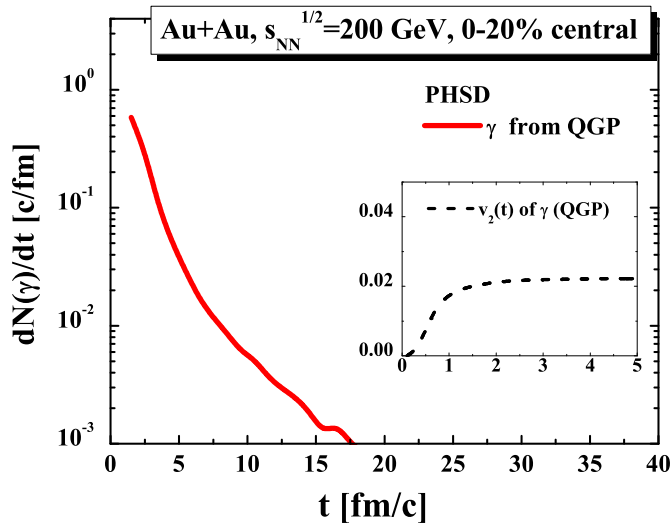}
\caption{(Color on-line) Rate of photon production in the QGP versus
time for 0-20\% Au+Au collisions at $\sqrt{s_{NN}}$ = 200 GeV from
the PHSD. The insert shows the time evolution of the elliptic flow
of photons from the partonic sources. \label{dNdt} }
\end{figure}

\begin{figure}
\includegraphics[width=0.45\textwidth]{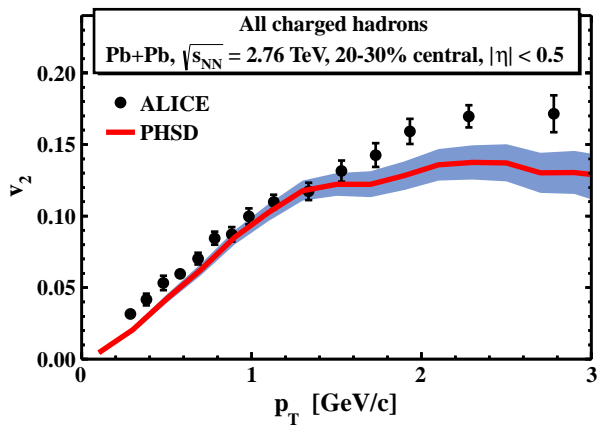}
\caption{(Color on-line) Elliptic flow $v_2$ versus transverse
momentum $p_T$ for the charged particles produced in  20-30\%
central Pb+Pb collisions at $\sqrt{s_{NN}}=2.76$~TeV from
the PHSD (solid red line) in comparison to the data from the ALICE
Collaboration \protect\cite{Aamodt:2010pa}; the blue band reflects
the statistical uncertainty of the PHSD calculations.
\label{chargedv2lhc} }
\end{figure}

More than a decade ago, the WA98 Collaboration has measured the
elliptic flow $v_2$ of photons produced in $Pb+Pb$ collisions at the
beam energy of $E_{beam}=158$~AGeV~\cite{Aggarwal:2004zh}, and it
was found that the $v_2(\gamma^{incl})$ of the
low-transverse-momentum inclusive photons was equal to the
$v_2(\gamma^{\pi})$ of pions within the experimental uncertainties.
This observation has lead to the conclusion that either (Scenario
1:)  the elliptic flow of the direct photons was comparable in
magnitude to the $v_2(\gamma^{incl})$ and $v_2(\gamma^{decay})$, or
(Scenario 2:) the contribution of the direct photons to the
inclusive ones was negligible.
However, the photon spectrum measured by the WA98 Collaboration
showed a significant yield of direct photons at low transverse
momentum. Thus the scenario 2 can be ruled out. Consequently, one
has to assume that the observed direct photons of low $p_T$ had a
significant elliptic anisotropy $v_2$ -- of the same order of
magnitude as the hadronic flow. The
interpretation~\cite{Liu:2007zzw,Bratkovskaya:2008iq} of the
low-$p_T$ direct photon yield measured by WA98 as dominantly
produced by the bremsstrahlung process in the mesonic collisions
$\pi+\pi\to\pi+\pi+\gamma$ is in accord with the WA98 data on the
inclusive photon $v_2(\gamma^{incl})$.

\begin{figure}
\includegraphics[width=0.45\textwidth]{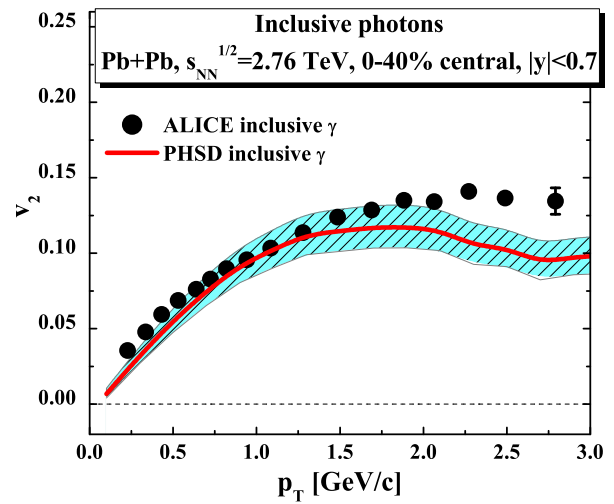}
\caption{(Color on-line) Elliptic flow $v_2$ versus transverse
momentum $p_T$ for the {\em inclusive} photons produced in  0-40\%
central Pb+Pb collisions at $\sqrt{s_{NN}}=2.76$~TeV from
the PHSD (solid red line) in comparison to the data from
the ALICE Collaboration \protect\cite{Lohner:2012ct}; the blue error
band reflects the finite statistics and the uncertainty in the
modeling of the cross sections for the individual channels.
\label{inclv2lhc} }
\end{figure}

Let us note that the same conclusions apply also to the most recent
studies of the photon elliptic flow at RHIC and LHC. The PHENIX and
ALICE Collaborations have measured the inclusive photon $v_2$ and
found that at low transverse momenta it is comparable to the
$v_2(p_T)$ of decay photons as calculated in cocktail simulations
based on the known mesonic $v_2(p_T)$. Therefore (a) either the
yield of the direct photons to the inclusive ones is not
statistically significant in comparison to the decay photons or (b)
the elliptic flow of the direct photons must be as large as
$v_2(\gamma^{decay})$ and $v_2(\gamma^{incl})$.

\begin{figure}
\includegraphics[width=0.45\textwidth]{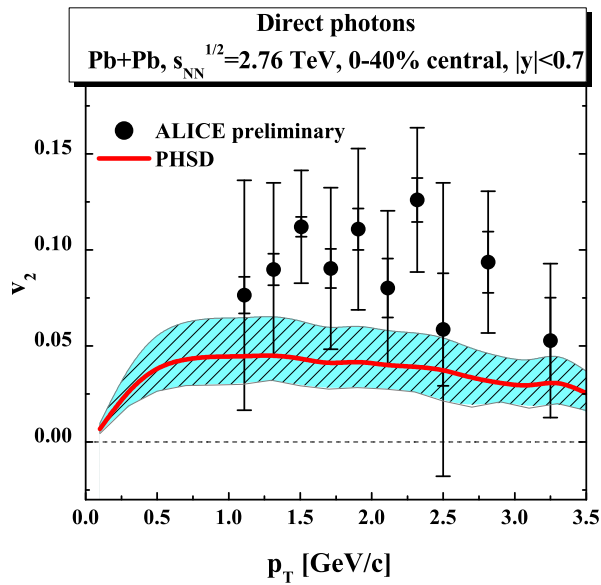}
\caption{(Color on-line) Elliptic flow $v_2$ versus transverse
momentum $p_T$ for the direct photons produced in  0-40\% central
Pb+Pb collisions at $\sqrt{s_{NN}}=2.76$~TeV from the
PHSD (solid red line) in comparison to the preliminary data from the
ALICE Collaboration \protect\cite{Lohner:2012ct}; the blue error
band is dominated by the uncertainty in the modeling of the cross
sections for the individual channels. \label{lhcv2dir} }
\end{figure}

In Refs.~\cite{Linnyk:2013hta,Linnyk:2013wma} we have shown the
elliptic flow of the inclusive and direct photons produced in
minimum bias Au+Au collisions at $\sqrt{s_{NN}}=200$~GeV from the
PHSD in comparison to the data of the PHENIX Collaboration. We found
that the data on the inclusive photon $v_2$ could approximately be
described. Furthermore, the pion decay photons dominate the
inclusive photon spectrum. Since the elliptic flow of pions is under
control in PHSD  in comparison to the data from the PHENIX and STAR
Collaborations (cf.
Refs.~\cite{Linnyk:2013hta,Konchakovski:2014fya,PHENIX1,Adams:2004bi,Adler:2003kt}),
the spectrum of decay photons is also predicted reliably by the
model. However, the good agreement with the inclusive photon
spectrum is especially meaningful due to the good description of the
direct photon spectrum which was presented above in
section~\ref{sect:results}.

After subtracting the contribution of the decay photons from the
inclusive photons, the direct photon $v_2$ is accessed
experimentally. In the PHSD, we calculate the direct photon
$v_2(\gamma^{dir})$ by building the weighted sum of the channels,
which are not subtracted by the data-driven methods, as follows: the
photons from the quark-gluon plasma, from the initial hard parton
collisions (pQCD photons), from the decays of short-living
resonances ($a_1$-meson, $\phi$-meson, $\Delta$-baryon), from the
binary meson+meson and meson+baryon channels
($\pi+\rho\to\pi+\gamma$, $\pi+\pi\to\rho+\gamma$, $V+p/n\to n /
p+\gamma$), and from the bremsstrahlung in the elastic meson+meson
and meson+baryon collisions ($m+m\to m+m+\gamma$, $m+B\to
m+B+\gamma$). We calculate the direct photon $v_2$ by summing up the
elliptic flow of the individual channels contributing to the direct
photons, using their contributions to the spectrum as the relative
$p_T$-dependent weights, $w_i(p_T)$, i.e.
\be \label{dir2}  v_2 (\gamma^{dir}) = \sum _i  v_2 (\gamma^{i}) w_i
(p_T) =  \frac{\sum _i  v_2 (\gamma^{i}) N_i (p_T)}{\sum_i N_i
(p_T)}. \ee

\begin{figure*} \centering
\includegraphics[width=0.9\textwidth]{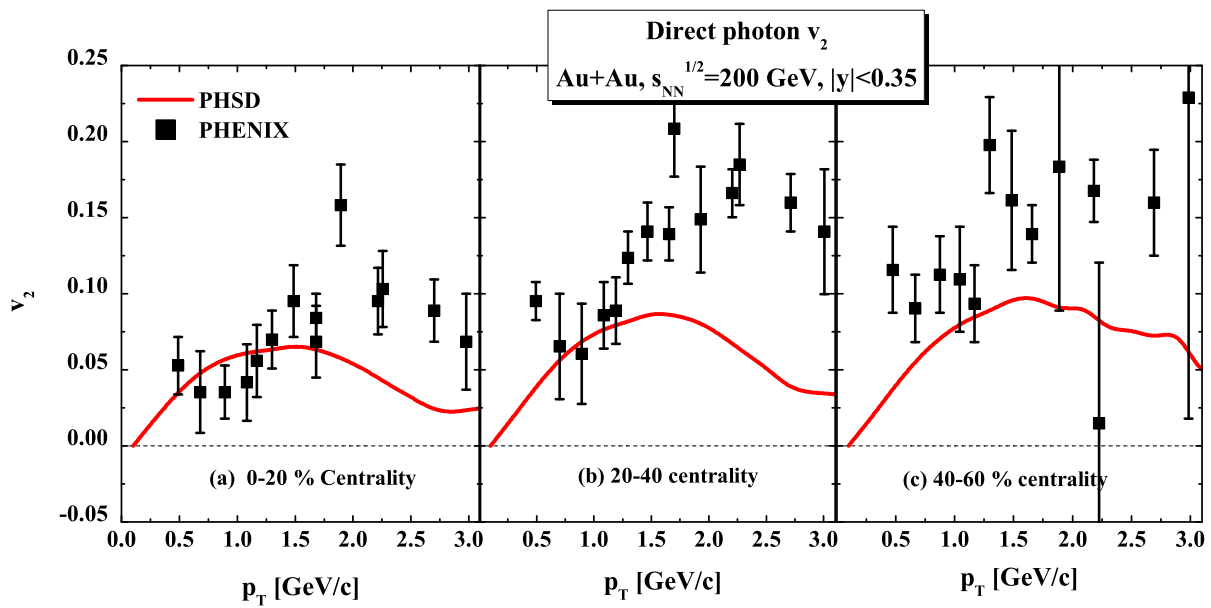}
\caption{ Centrality dependence of the direct photon $v_2$ for Au+Au
collisions at $\sqrt{s_{NN}}$ = 200 GeV for different centralities
(see legend); the data from the PHENIX
Collaboration~\protect\cite{Adare:2014fwh,Adare:2015lcd} are
compared to the earlier PHSD predictions from
Ref.~\protect\cite{Linnyk:2013wma}. } \label{v2b1}
\end{figure*}

The results for the elliptic flow $v_2(p_T)$ of direct photons
produced in $Au+Au$ collisions at the top RHIC energy are shown in
Fig.~\ref{rhicV2}.  In comparison to the previous results within the
PHSD approach~\cite{Linnyk:2013hta,Linnyk:2013wma}, the elliptic
flow in the intermediate region of the transverse momenta
$1.0<p_T<2.0$~GeV is reduced by about 50\% due to the modifications
in our computation of the bremsstrahlung channels beyond the
soft-photon approximation in the present work. According to our
calculations of the direct photon spectra (as presented above),
almost half of the direct photons measured by PHENIX (in central
collisions) stems from the collisions of quarks and gluons in the
deconfined medium created in the initial phase of the collision. The
photons produced in the QGP carry a very small $v_2$ and lead to an
overall direct photon $v_2$ about a factor of 2 below the pion
$v_2(\pi)$ even though the other channels in the sum (\ref{dir2})
have large elliptic flow coefficients $v_2$ of the order of
$v_2(\pi)$ (cf. Fig.~7 of Ref.~\cite{Linnyk:2013hta}).

\begin{figure}
\includegraphics[width=0.45\textwidth]{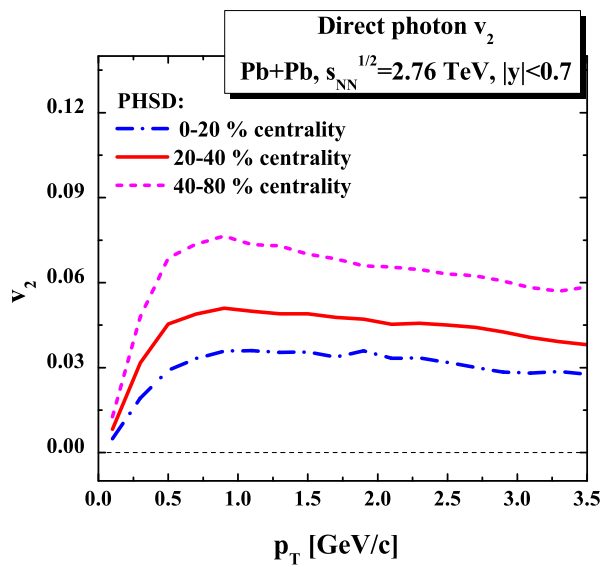}
\caption{(Color on-line) Elliptic flow $v_2$ of direct photons from PHSD versus transverse momentum $p_T$
produced in
Pb+Pb collisions at $\sqrt{s_{NN}}$ = 2.76 TeV for different centrality classes (see legend).
\label{v2b2} }
\end{figure}

Indeed, the parton collisions -- producing photons in the QGP --
take place throughout the evolution of the collision but the
collision rate falls rapidly with time and thus the production of
photons from the QGP is dominated by the early times.  We illustrate
this in Fig.\ref{dNdt}, where it is seen that the rate of photon
emission $dN_\gamma/dt$ from QGP sources drops by orders of magnitude after the first 10
fm/c. As a consequence, the elliptic flow `picked up' by the photons
from the parent parton collisions saturates after about a few fm/c and
reaches a relatively low value of about $0.02$, only. We note that a
delayed production of charges from the strong gluon fields
(`glasma'~\cite{McLerran:2011zz,Chiu:2012ij,Blaizot:2012qd,Gale:2012rq})
might shift the QGP photon production to somewhat later times when
the elliptic flow is built up more. However, we cannot
quantitatively ans\-wer whether the additional evolution in the
pre-plasma state could generate considerable additional direct photon $v_2$.

In the following we  present calculations/predictions for the elliptic flow of
inclusive and direct photons produced in  $Pb+Pb$ collisions at the
energy of $\sqrt{s_{NN}}=2.76$~TeV.
First, we show in Fig.~\ref{chargedv2lhc} the elliptic flow $v_2$
versus transverse momentum $p_T$ for the charged hadrons produced
in  20-30\% central Pb+Pb collisions at $\sqrt{s_{NN}}=2.76$~TeV
from the PHSD (solid red line); the blue band reflects the
statistical uncertainty of the calculations. The PHSD results are
compared to the data of the ALICE Collaboration \cite{Aamodt:2010pa} which suggests that
the bulk dynamics is reasonably under control in the PHSD at these
energies and the elliptic flow of the final charged particles is described
up to  $p_T \approx 2$~GeV/c in this centrality range.

We, furthermore, present our calculations/predictions for the
elliptic flow of inclusive and direct photons produced in  $Pb+Pb$
collisions at the energy of $\sqrt{s_{NN}}=2.76$~TeV at the LHC
within the acceptance of the ALICE detector. Fig.~\ref{inclv2lhc}
shows PHSD calculations for the elliptic flow $v_2$ versus
transverse momentum $p_T$ for the {\em inclusive} photons produced
in 0-40\% central Pb+Pb collisions at $\sqrt{s_{NN}}=2.76$~TeV
(solid red line) with the blue error band reflecting the finite
statistics and the theoretical uncertainty in the modeling of the
cross sections. A comparison to the respective data from the ALICE
Collaboration \cite{Lohner:2012ct} shows a comparable agreement as
in case of the $v_2$ for the charged hadrons.

The elliptic flow $v_2(p_T)$ of {\em direct} photons produced in
0-40\% central Pb+Pb collisions at $\sqrt{s_{NN}}=2.76$~TeV from
the PHSD (solid red line) is shown in
Fig.~\ref{lhcv2dir} in comparison to the preliminary data from the
ALICE Collaboration \cite{Lohner:2012ct}; the blue error band is
again dominated by the uncertainty in the modeling of the cross
sections for the individual channels. As in case of Au+Au collisions
at the top RHIC energy we slightly underestimate the direct photon
$v_2$, however, new data with higher accuracy will be needed to shed
further light on the direct photon $v_2$ puzzle.

The centrality dependence of the direct photon elliptic flow has
been calculated within the PHSD in Ref.~\cite{Linnyk:2013wma} for Au+Au
collisions at the top RHIC energy.
Fig. ~\ref{v2b1} presents a direct comparison of the PHSD predictions for $v_2(p_T)$
from Ref.~\cite{Linnyk:2013wma} for  Au+Au collisions at
$\sqrt{s_{NN}}$=200 GeV in the centrality classes 0-20\% (a), 20-40\% (b) and 40-60\% (c) with
the data from Refs. ~\cite{Adare:2008ab,Adare:2014fwh}.
Whereas the elliptic flow is roughly described in the most central class there is an increasing tendency
to underestimate in the PHSD the strong elliptic flow especially for peripheral
collisions where some additional source might be present. Thus the observed
centrality dependence of the elliptic flow is roughly in agreement with the
interpretation that a large fraction of the direct photons is of hadronic origin (in
particular from the bremsstrahlung in meson+meson and meson+baryon
collisions); the latter contribution becomes stronger in more peripheral collisions.
But more precise data will be mandatory for a robust conclusion.

We close this Subsection in providing predictions for the centrality dependence
of the direct photon $v_2(p_T)$ in Pb+Pb collisions at $\sqrt{s_{NN}}$ = 2.76 TeV in the  centrality classes
0-20\%, 20-40\% and 40-80\%  which are of
relevance for the upcoming measurements by the ALICE Collaboration at
the LHC. The actual results from PHSD are displayed in Fig. \ref{v2b2} and show a very similar centrality dependence
as in case of Au+Au collisions at the top RHIC energy.

\begin{figure*}
\includegraphics[width=0.9\textwidth]{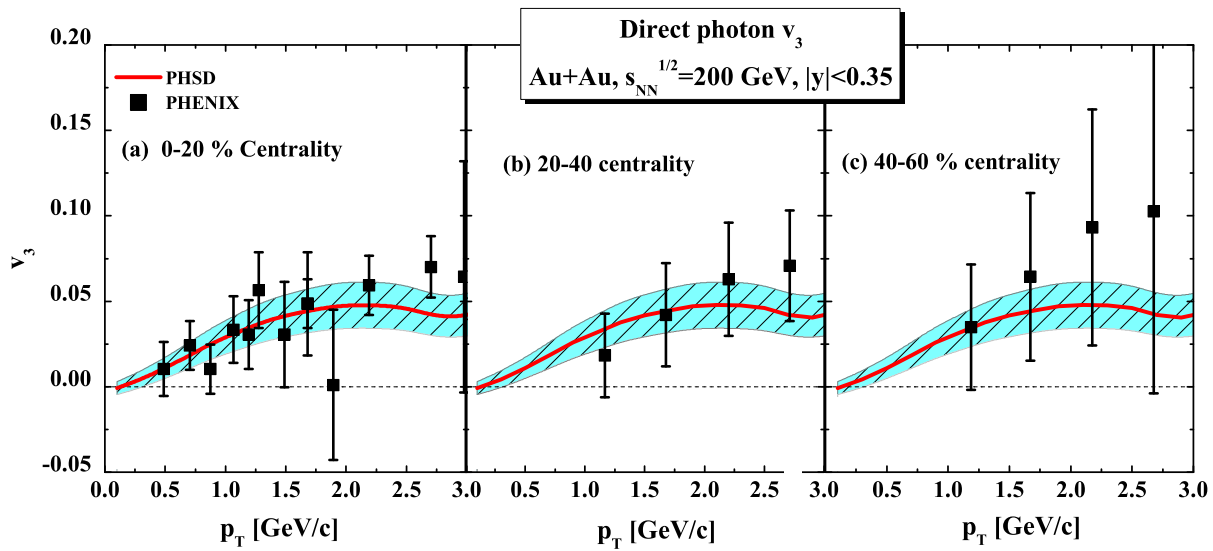}
\caption{(Color on-line) Triangular flow $v_3$ versus transverse
momentum $p_T$ for the direct photons produced in Au+Au collisions
at $\sqrt{s_{NN}}=200$~GeV in three centrality classes (see
legends).
The PHSD results are shown by the solid red lines in comparison to
the data of the PHENIX Collaboration (black symbols) taken from
Ref.~\protect{\cite{Bannier:2014bja,Adare:2015lcd}}.} \label{v3}
\end{figure*}

\subsection{Triangular flow of direct photons}
\label{sect:v3}

We have seen in the previous sections that the measured spectra of
direct photons could be reproduced by the PHSD calculations within a
factor of 2 (which is comparable with the current accuracy of the
measurements). Also, the {\em inclusive} photon $v_2$ was well described  and
the elliptic flow of {\em direct} photons was qualitatively in line with
the data (within a factor of 2) and attributed essentially to hadronic sources.

On the other hand, there exists an alternative interpretation of the
strong elliptic flow of direct photons, in which the azimuthal
asymmetry of the photons is due to the initial strong magnetic field
essentially produced by spectator charges (protons). Indeed, the
magnetic field strength in the very early reaction stage reaches up
to $eB_y \approx 5 m_\pi^2$ in semi-peripheral $Au+Au$ collisions at
$\sqrt{s_{NN}}=200$~GeV (see the calculations within the PHSD in
Ref.~\cite{Voronyuk:2011jd}; comparable estimates have been obtained
also in Refs.~\cite{Tuchin:2014pkaTuchin:2012mf,Skokov:2009qp}).
These strong magnetic fields might influence the photon production
via the polarization of the medium, e.g. by influencing the motion
of charged quarks in the QGP, or by directly inducing a real photon
radiation via the virtual photon ($\vec B$-field) coup\-ling to a
quark loop and (multiple) gluons; the photons are then produced
azimuthal asymmetrically.

\begin{figure}
\includegraphics[width=0.45\textwidth]{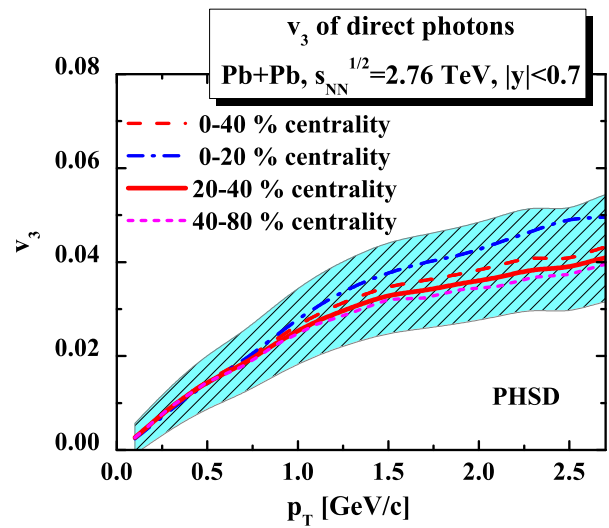}
\caption{(Color on-line) Triangular flow $v_3$ versus transverse
momentum $p_T$ for the direct photons produced in  different centrality classes for
Pb+Pb collisions at $\sqrt{s_{NN}}=2.76$~TeV from the
PHSD (see legend); the blue band reflects the uncertainty in the
modeling of the cross sections for the individual channels and give a
measure of the present level of accuracy.
\label{v3_lhc}}
\end{figure}

The photon production under the influence of strong magnetic fields has
been calculated in
Refs.~\cite{Bzdak:2012fr,Tuchin:2014pkaTuchin:2012mf,McLerran:2014hza,Skokov:2013axaBasar:2012bp}.
The observed spectra and elliptic flow of direct photons could be
explained using suitable assumptions on the conductivity, bulk
viscosity or degree of chemical equilibration in the early produced
matter. The common feature of these calculations was that the {\em
triangular} flow coefficient $v_3$ of the direct photons was
expected to be very small. Indeed, the magnetic field may lead to an
azimuthal asymmetry $v_2$ but not to a triangular mode.

Consequently, it is of interest to measure experimentally the third
flow coefficient $v_3(p_T)$ and to compare it to the calculations in
the different classes of models: (a) those attributing the large
elliptic flow and strong yield of direct photons to hadronic
sources, e.g. the PHSD transport approach; (b) the models suggesting
the large azimuthal asymmetry and additional yield of direct photons
to be caused by the early magnetic fields; (c) the models assuming
that the yield of direct photons at low $p_T$ is dominated by
partonic channels.

In Fig.~\ref{v3} we present our results for the triangular flow
$v_3$ versus transverse momentum $p_T$ for the direct photons
produced in Au+Au collisions at $\sqrt{s_{NN}}=200$~GeV from
the PHSD (solid red lines) for 0-20\% (a), 20-40\% (b) and 40-60\% (c) centrality.
The PHSD gives a positive non-zero triangular flow of direct photons
up to 6\% with very little centrality dependence on the level of the present accuracy ($\sim 25\%$).
The PHSD results are in agreement
with the data of the PHENIX Collaboration from
Refs.~\cite{Bannier:2014bja,Mizuno:2014via,Ruan:2014kia} which
suggests that the scenario (a) is compatible with the measurements.

Accordingly, the  data from PHENIX in Fig. \ref{v3}  do not point towards an interpretation
of the direct photons being dominantly produced in the early stage
under the influence of the magnetic field (b), because the $v_3$ of
these photons is expected to be close to zero. Of course, the photon
production in the magnetic fields occurs on top of other channels,
which may carry finite $v_3$. But the weighted sum of all the
channels including the magnetic-field-induced photons will give a
smaller $v_3\ne0$  than the sum without this channel. The scenario
(c) has been studied by other groups within a hydrodynamic modeling
of the collision  in Refs.~\cite{Shen:2013cca,Chatterjee:2014dqa}.
The triangular flow $v_3(p_T)$ of direct photons from
Refs.~\cite{Shen:2013cca,Chatterjee:2014dqa} is about a factor of 2
smaller than that obtained in the present work from the PHSD. It will be possible
to differentiate between the scenarios in the future, when  data of
higher accuracy will become available.

We finally show in Fig.~\ref{v3_lhc} our predictions for the triangular flow
$v_3(p_T)$ of direct photons produced in
$Pb+Pb$ collisions at $\sqrt{s_{NN}}=2.76$~TeV for different centralities as will be measured
at the LHC. We expect the triangular flow of direct photons to reach
at maximum $v_3^{max}=0.04\pm0.015$, thus being comparable to the
charged hadron $v_3$~\cite{Konchakovski:2014fya}. The centrality dependence of $v_3(p_T)$
turns out to be low and is practically constant within the accuracy of
the PHSD calculations. An experimental  confirmation of
this expectation could further affirm the notion of large
hadronic contributions to the direct photons and in particular the
photon production via the bremsstrahlung in meson and baryon
collisions.

\section{Summary}
\label{sect:summary}

In the present work we have calculated the transverse momentum spectra, the
elliptic flow $v_2$ and triangular flow $v_3$ of direct photons
produced in Au+Au collisions at $\sqrt{s_{NN}}=200$~GeV and in
$Pb+Pb$ collisions at $\sqrt{s_{NN}}=2.76$~TeV using the microscopic
PHSD transport approach. For thermal photon production we have
considered the interactions of quarks and gluons in the strongly
interacting quark-gluon plasma (sQGP) ($q+\bar q\to g+\gamma$ and
 $q(\bar q)+g\to q(\bar q)+\gamma$), the photon production in the hadronic decays
($\pi\to\gamma+\gamma$, $\eta\to\gamma+\gamma$,
$\omega\to\pi+\gamma$, $\eta'\to\rho+\gamma$, $\phi\to\eta+\gamma$,
$a_1\to\pi+\gamma$, $\Delta\to N+\gamma$) as well as the
interactions ($\pi+\pi\to\rho+\gamma$, $V+\pi\to\pi+\gamma$, the
$2\to2$ interaction of mesons and baryons $\rho+n/p\to\gamma+n/p$,
and last but not least the bremsstrahlung radiation in meson+meson
and meson+baryon scattering $m+m/B\to m+m/B+\gamma$ throughout the
evolution of the collision. The pQCD photons produced in the initial
hard binary scatterings are added to the thermal photons in order to
obtain the final direct photon spectrum.

In extension of our previous works on photon production in heavy-ion
collisions  at the top RHIC energy we \\
(i) went beyond the
soft-photon approximation (SPA) in the calculation of the
bremsstrahlung processes $meson+meson\to meson+meson+\gamma$,
$meson+baryon\to meson+baryon+\gamma$, \\
(ii) quantified the
suppression at low $p_T$ due the Landau-Pomeranchuk-Migdal (LPM)
effect in connection with the electric conductivity, and \\
(iii)
incorporated the $V+N\to\gamma+N$ and $\Delta\to N+\gamma$ channels
into the PHSD. Furthermore, \\
(iv)  have presented calculations for
Pb+Pb collisions at SPS and LHC energies.

The result from Ref.~\cite{Liu:2007zzw} for the reaction $\pi+\pi
\rightarrow \pi + \pi + \gamma$, obtained within the one-boson
exchange model {\em beyond} the soft-photon approximation
 -- available up to the
photon energy of 0.4 GeV -- is confirmed by our present calculations
that extend to 2 GeV. Furthermore, the {\em improved} SPA
(\ref{int_spa_formula2}) gives a very good approximation to the
exact result even at high $\sqrt{s}$ as long as the model for the
elastic cross section is sufficiently realistic. In comparison, the
constant-cross-section approximation based on the formula
(\ref{brems}) (used before) overestimates the exact rates for
$q_0>1$~GeV and underestimates for $q_0<0.4$~GeV. The OBE
differential cross sections have been implemented in the PHSD
transport approach for photon production studies
in heavy-ion reactions from SPS to LHC energies.

Furthermore, we have given an estimate for the photon suppression
due to the Landau-Pomeranchuk-Migdal (LPM) effect employing the approximation
(\ref{lenk}) that captures the relevant physics. When inserting the interaction time
of the partons from PHSD at finite temperature we obtain a photon rate that is well in line
with Eq.  (\ref{ratecond}) in the limit of vanishing photon energy.
We recall that the electric conductivity $\sigma_0$
regulates the divergence of the photon rate and has been
evaluated in the PHSD in Ref. \cite{Cassing:2013iz} in rough agreement with related results from lattice QCD.
The actual results in Fig. 9 demonstrate that the LPM effect becomes important for photon energies
below 0.4 GeV in case of dense and strongly interacting partons. Since the experimental photon spectra
measured so far at RHIC and LHC energies start from higher phton energies the LPM effect
might be safely discarded.

In case of relativistic heavy-ion collisions
we have found that the PHSD calculations reproduce the transverse
momentum spectra of direct photons as measured by the PHENIX
Collaboration in Refs.~\cite{PHENIXlast,Adare:2008ab} for Au+Au
collisions at the top RHIC energy. The calculations reveal the
channel decomposition of the observed direct photon spectrum and
show that the photons produced in the QGP constitute at most about
50\% of the direct photons in central collisions with the rest being
distributed among the other channels: mesonic interactions, decays
of massive hadronic resonances and the initial hard scatterings.
Our calculations demonstrate that the photon production in the QGP
is dominated by the early phase (similar to hydrodynamic models) and
is localized in the center of the fireball, where the collective
flow is still rather low, i.e. on the 2-3 \% level, only.
Thus, the strong $v_2$ of direct photons - which is comparable to
the hadronic $v_2$ - in PHSD is attributed to hadronic channels,
i.e. to meson and baryon induced reactions. On the other hand, the
strong $v_2$ of the 'parent' hadrons, in turn, stems from the
interactions in the QGP via collisions and the partonic mean-field
potentials. Accordingly, the presence of the QGP shows up
'indirectly' in the direct photon elliptic flow.

We, furthermore, have demonstrated that the elliptic flow of charged
hadrons from PHSD is in a reasonable agreement with the ALICE
data. Predictions/calculations for the {\em inclusive} photon $v_2$ and {\em direct} photon
$v_2$ have been provided as well as associated results for the
triangular direct photon flow $v_3(p_T)$ which is as large as the
charged hadron $v_3$ at $\sqrt{s_{NN}}$ = 200 GeV and
$\sqrt{s_{NN}}$ = 2.76 TeV. The large triangular flow of direct
photons seen by PHENIX at the top RHIC energy is in line with the
PHSD calculations and in conflict with a scenario that attributes
the large direct photon $v_2$ to the influence of the early strong
magnetic field from the spectator charges.

The centrality dependence of the direct photon yield and flow
has the potential to further clarify the
direct photon production mechanisms. We find a good agreement
between the PHENIX measurements and the PHSD calculations at the top RHIC energy. In
particular, the integrated thermal photon yield in PHSD was
predicted to scale as $N_{part}^\alpha$ with the exponent
$\alpha \approx 1.5$, which is in a good agreement with the most recent
measurement of $\alpha=1.48\pm0.08\pm0.04$ by the PHENIX
Collaboration~\cite{Adare:2014fwh}. This observation supports the
conclusion that the low transverse momentum direct photons have a
strong contribution from the binary hadronic photon production
sources, such as the $meson+meson$ and $meson+baryon$
bremsstrahlung. It will be important to investigate experimentally
the scaling of the direct photon yield and flows $v_2$ and $v_3$ with centrality also
in Pb+Pb collisions at the LHC energies for which we have provided explicit predictions from the PHSD.


\section*{Acknowledgements}

We appreciate fruitful discussions with C.~Gale, J. F.~Paquet,
C.~M.~Ko, D.~Cabrera, H.~Berrehrah, L.~McLerran, K.~Eskola,
I.~Tserruya, C.~Klein-Boesing, R.~Rapp, H.~van Hess, J.~Stachel,
U.~Heinz, I.~Selyuzhenkov, G.~David, K.~Redlich, K.~Reygers,
C.~Shen, A.~Drees, B.~Bannier and F.~Bock. The contributions of
J. F.~Paquet, C.~Gale, K.~Eskola, I.~Helenius and C.~Klein-Boesing
have been essential for constraining the pQCD photon yield. We especially
thank C.~Gale for enriching this work with his ideas and
support during his stay at the Frankfurt Institute for Advanced
Studies. Furthermore, the authors are grateful to K.~Reygers,
J.~Stachel and the ExtreMe Matter Institute for conducting a
stimulating and productive EMMI-Rapid-Reaction-Task-Force meeting on
the ``Direct-Photon Flow Puzzle". Furthermore, we acknowledge
financial support through the ``HIC for FAIR" framework of the
``LOEWE" program. The computational resources have been provided by the
LOEWE-CSC as well as the SKYLLA cluster at the Univ. of Giessen.




\end{document}